\documentclass[10pt,twocolumn,letterpaper]{article}

\usepackage[pagenumbers]{cvpr} 

\makeatletter
\@namedef{ver@everyshi.sty}{}
\makeatother

\usepackage{standalone}

\usepackage{mathtools} 

\usepackage{graphicx}
\usepackage{amsmath}
\usepackage{amssymb}
\usepackage{amsthm}
\usepackage{booktabs}
\usepackage[linesnumbered,ruled]{algorithm2e}

\SetCommentSty{mycommfont}
\usepackage{enumerate}
\usepackage{twoopt}
\usepackage{tikz}
\usetikzlibrary{positioning}
\usetikzlibrary{snakes}
\usetikzlibrary{decorations.pathmorphing}
\usetikzlibrary{calc}
\usetikzlibrary{arrows}
\usetikzlibrary{decorations.markings}
\usetikzlibrary{fit}
\usepackage{microtype}
\usepackage{pgfplots}
\usepackage{color}
\usepackage[accsupp]{axessibility}  

\definecolor{mycolor1}{HTML}{009900}
\definecolor{mycolor2}{HTML}{FF0000}

\definecolor{scattercolor1}{HTML}{800080}
\definecolor{scattercolor2}{HTML}{990000}
\definecolor{scattercolor3}{HTML}{009900}
\definecolor{scattercolor4}{HTML}{808000}
\definecolor{scattercolor5}{HTML}{D30E78}
\definecolor{scattercolor6}{HTML}{036ffc}
\definecolor{scattercolor7}{HTML}{eb9800}

\definecolor{green}{HTML}{00FF00}
\definecolor{red}{HTML}{FF0000}

\pgfplotsset{compat=newest}
\pgfplotsset{/pgfplots/error bars/error bar style={thick}}
\pgfplotsset{/pgfplots/error bars/error mark options={line width=1pt, mark size=3pt, rotate=90}}

\newcommand{\R}{\mathbb{R}}

\newcommand{\N}{\mathbb{N}}
\newcommand{\M}{\mathcal{M}}

\newcommand{\diag}{\operatorname{diag}}

\DeclareMathOperator*{\argmin}{arg\,min}
\DeclarePairedDelimiter\abs{\lvert}{\rvert}%
\DeclarePairedDelimiter\norm{\lVert}{\rVert}%
\newcommand{\la}{\langle}
\newcommand{\ra}{\rangle}
\newcommand{\conv}{\mathsf{conv}}

\DeclareMathOperator{\NE}{\mathcal{N}}

\DeclareMathOperator{\OO}{\mathcal{O}}
\DeclareMathOperator{\HH}{\mathcal{H}}
\DeclareMathOperator{\HHte}{\mathcal{H}_{T \rightarrow E}}
\DeclareMathOperator{\HHet}{\mathcal{H}_{E \rightarrow T}}

\let\oldnl\nl
\newcommand{\nonl}{\renewcommand{\nl}{\let\nl\oldnl}}

\def\myparagraph#1{\vspace*{3.5pt}\noindent{\bf #1~~}}
\usepackage{enumitem} 
\newenvironment{packed_description}{
\begin{description}[leftmargin=12pt]
  \setlength{\itemsep}{3pt}
  \setlength{\parskip}{0pt}
  \setlength{\parsep}{0pt}
}{\end{description}}
\newtheorem{theorem}{Theorem}
\newtheorem{lemma}[theorem]{Lemma}

\newtheorem{definition}[theorem]{Definition}

\usepackage{hyphenat}
\usepackage[pagebackref,breaklinks,colorlinks]{hyperref}

\usepackage[capitalize]{cleveref}
\crefname{section}{Sec.}{Secs.}
\Crefname{section}{Section}{Sections}
\Crefname{table}{Table}{Tables}
\crefname{table}{Tab.}{Tabs.}

\def\cvprPaperID{1089} 
\def\confName{CVPR}
\def\confYear{2022}

\begin{document}

\title{RAMA: A Rapid Multicut Algorithm on GPU}

\author{Ahmed Abbas \qquad Paul Swoboda \\
\small{Max Planck Institute for Informatics, Saarland Informatics Campus}

}
\maketitle

\begin{abstract}
We propose a highly parallel primal-dual algorithm for the multicut (a.k.a.\ correlation clustering) problem, a classical graph clustering problem widely used in machine learning and computer vision.
Our algorithm consists of three steps executed recursively:
(1) Finding conflicted cycles that correspond to violated inequalities of the underlying multicut relaxation,
(2) Performing message passing between the edges and cycles to optimize the Lagrange relaxation coming from the found violated cycles producing reduced costs and
(3) Contracting edges with high reduced costs through matrix-matrix multiplications.

Our algorithm produces primal solutions and lower bounds that estimate the distance to optimum.
We implement our algorithm on GPUs and show resulting one to two orders-of-magnitudes improvements in execution speed without sacrificing solution quality compared to traditional sequential algorithms that run on CPUs.
We can solve very large scale benchmark problems with up to  $\mathcal{O}(10^8)$ variables in a few seconds with small primal-dual gaps. Our code is available at~\url{https://github.com/pawelswoboda/RAMA}.
\end{abstract}
\section{Introduction}
\label{sec:introduction}

Decomposing a graph into meaningful clusters is a fundamental problem in combinatorial optimization.
The multicut problem~\cite{chopra1993partition} (also known as correlation clustering~\cite{bansal2004correlation}) is a popular approach to decompose a graph into an arbitrary number of clusters based on affinites between nodes.

The multicut problem and its extensions such as higher order multicut~\cite{kim2011higher,kappes2016higher}, lifted multicut~\cite{keuper2015efficient}, (asymmetric) multiway cut~\cite{chopra1991multiway,kroeger2014asymmetric}, lifted disjoint paths~\cite{hornakova2020lifted} and joint multicut and node labeling~\cite{levinkov2017joint} have found numerous applications in machine learning, computer vision, biomedical image analysis, data mining and beyond.
Examples include 
unsupervised image segmentation~\cite{alush2013break,andres2011closedness,yarkony2012fast,andres2013segmenting}, 
instance-separating semantic segmentation\cite{kirillov2017instancecut,abbas2021combinatorial},
multiple object tracking~\cite{tang2017multiple,hornakova2020lifted},
cell tracking~\cite{jug2016moral},
articulated human body pose estimation~\cite{insafutdinov2017art},
motion segmentation~\cite{keuper2018motion},
image and mesh segmentation~\cite{keuper2015efficient},
connectomics~\cite{andres2012globally,beier2017multicut,pape2017solving}
and many more.

Multicut and its extensions are NP-hard to solve~\cite{bansal2004correlation,demaine2006correlation}.
Since large problem instances with millions or even billions of variables typically occur, powerful approximative algorithms have been developed~\cite{keuper2015efficient,swoboda2017dual,beier2014cut,beier2016efficient,levinkov2019comparative}.
However, even simple heuristics such as~\cite{keuper2015efficient} require very large running times for very large instances.
In particular, some instances, such as those investigated in~\cite{pape2017solving} could not be solved in acceptable time (hence ad-hoc decomposition techniques were used).
In other scenarios very fast running times are essential, \eg when multicut is used in end-to-end training~\cite{song2019end,abbas2021combinatorial}.
Hence, the need for parallelization arises, preferably on GPUs.
The parallelism offered by GPUs is typically difficult to exploit due to irregular data structures and the inherently sequential nature of most combinatorial optimization algorithms.
This makes design of combinatorial optimization algorithms challenging on GPUs.
An additional benefit of running our algorithms on GPU is that memory transfers between CPU and GPU are avoided when used in a deep learning pipeline.

Our contribution is a new primal-dual method that can be massively parallelized and run on GPU.
This results in faster runtimes than previous multicut solvers while still computing solutions which are similar or better than CPU based solvers in terms of objective.
Yet, our approach is rooted in solving a principled polyhedral relaxation and yields both a primal solution and a dual lower bound.
In particular, finding primal solutions and approximate dual solving is interleaved such that both components of our algorithm can profit from each other.
In more detail, our algorithmic contribution can be categorized as follows

\begin{packed_description}
\item[\textit{Primal:} Edge Contraction:]
Finding a primal solution depends similarly as in~\cite{keuper2015efficient} on contracting edges that are highly likely to end up in the same component of the final clustering.
To this end we propose to use a linear algebra approach by expressing edge contractions as sparse matrix-matrix multiplications.
This allows us to accelerate edge contraction by exploiting highly parallel matrix-matrix multiplication GPU primitives.

\item[\textit{Dual:} Lagrange Relaxation \& Message Passing:]
To find good edge contraction candidates, we consider approximately solving a polyhedral relaxation by searching for conflicting cycles, adding them to a Lagrange relaxation and updating the resulting Lagrange multipliers iteratively by message passing.
We propose a new message passing scheme that is both massively parallelizable yet yields monotonic increases w.r.t.\ the dual objective, speeding up the scheme of~\cite{swoboda2017dual} by orders of magnitude.

\item[\textit{Recursive} Primal-Dual:]
We interleave the above operations of finding and solving a Lagrange relaxation and contracting edges, yielding the final graph decomposition.
Hence, our algorithm goes beyond classical polyhedral approaches~\cite{swoboda2017dual,kappes2011globally,nowozin2009solution} that only consider the original graph.
\end{packed_description}

On the experimental side we obtain primal solutions that are of comparable or better quality to those obtained by established high-quality heuristics~\cite{keuper2015efficient,lange2018partial} in a fraction of the execution time but with additional dual lower bounds that help in estimating the quality of the solutions.
We perform experiments on 2D and 3D instance segmentation problems for scene understanding~\cite{cordts2016cityscapes} and connectomics~\cite{pape2017solving} containing up to $\mathcal{O}(10^8)$ variables.
\section{Related Work}
\label{sec:related-work}
\myparagraph{Preprocessing and Inprocessing:}
For fixing variables to their optimal values and shrinking the problem before or during optimization, persistency or partial optimality methods have been proposed in~\cite{alush2012ensemble,lange2018partial,lange2019combinatorial}.
These methods apply a family of criteria that, when passed, prove that any solution can be improved if its values do not coincide with the persistently fixed variables.

\myparagraph{Primal heuristics:}
For obtaining primal solutions without optimality guarantees or estimates on the distance to optimum, a large number of methods have been proposed with different execution time/solution quality trade-offs. 
The first heuristic for multicut, the classical Kernighan\&Lin move-making algorithm was originally proposed in~\cite{kernighan1970efficient} and slightly generalized in~\cite{keuper2015efficient}. 
The algorithm consists of trying various moves such as joining two components, moving a node from one component to the next \etc and performing sequences of moves that decrease the objective.
The faster but simpler greedy additive edge contraction (GAEC) heuristic, a move making algorithm that only can join individual components, was proposed in~\cite{keuper2015efficient}.
It is used in~\cite{keuper2015efficient} to initialize the more complex Kernighan\&Lin algorithm.
Variants involving different join selection strategies were proposed in~\cite{kardoost2018solving}.
The greedy edge fixation algorithm~\cite{keuper2015efficient} generalizes GAEC in that it can additionally mark edges as cut, constraining their endpoints to be in distinct components.
The more involved Cut, Glue \& Cut (CGC) move-making heuristic~\cite{beier2014cut} works by alternating bipartitioning of the graph and exchanging nodes in pairs of clusters. 
The latter operation is performed by computing a max-cut on a planar subgraph via reduction to perfect matching.
CGC was extended to a more general class of possible ``fusion moves'' in~\cite{beier2016efficient}.
A parallel algorithm for the simpler problem of unweighted correlation clustering problem was given in~\cite{pan2015parallel}.
A comparative survey of some of the above primal heuristics is given in~\cite{levinkov2017comparative}.

\myparagraph{LP-based algorithms:}
For obtaining dual lower bounds that estimate the distance to the optimum or even certify optimality of a solution a number of LP-relaxation based algorithms have been proposed.
These algorithms can be used inside branch and bound and their computational results can be used to guide primal heuristics to provide increasingly better solutions.
Quite surprisingly, it has been shown by~\cite{kappes2011globally,kim2011higher} that multicut problems of moderately large sizes can be solved with commercial integer linear programming (ILP) solvers like Gurobi~\cite{gurobi} in a cutting plane framework in reasonable time to global optimality.
Column generation based on solving perfect matching subproblems has been proposed in~\cite{yarkony2012fast,lukasik2020benders}.
Still, the above approaches break down when truly large scale problems need to be solved, since the underlying LP-relaxations are still solved by traditional LP-solvers that do not scale linearly with problem size and are not explicitly adapted to the multicut problem.
Additionally, violated inequality separation (cutting planes) requires solving weighted shortest path problems which is not possible in linear time.
The message passing algorithm~\cite{swoboda2017message} approximately solves a dual LP-relaxation faster than traditional LP-solvers and has faster separation routines than those of primal LP-solvers as well, thereby scaling to larger problems.
An even faster, but less powerful, approximate cycle packing algorithm was proposed in~\cite{lange2018partial}.

\myparagraph{Other efficient clustering Methods:}
The mutex watershed~\cite{wolf2020mutex} and its generalizations~\cite{bailoni2019generalized} are closely related to the greedy additive edge fixation heuristic for multicut~\cite{levinkov2017comparative}.
The corresponding algorithms can be executed faster than their multicut counterparts on CPU, but are sequential.
Fast GPU schemes~\cite{auer2012graph} were proposed for agglomerative clustering.
Last, spectral clustering can be implemented on GPU with runtime gains~\cite{jin2016,naumov2016parallel}.
All these approaches however are not based on any energy minimization problem, hence do not come with the theoretical benefits that an optimization formulation offers.

\section{Method}
\label{sec:method}

\begin{figure}
\centering
    \tikzstyle{cut-edge}=[red,dashed,thick]
\tikzstyle{vertex}=[circle, draw, fill=white, inner sep=0pt, minimum width=1ex]
\tikzset{every picture/.append style={baseline,scale=1.1}}

\begin{tikzpicture}[scale=0.95]
\draw[draw=mycolor1!30, fill=mycolor1!30] plot[smooth cycle, tension=0.5] coordinates
{(-0.3, 2.3) (2.3, 2.3) (2.3, 0.7) (0.7, 0.7) (0.7, 1.7) (-0.3, 1.7)};
\draw[draw=mycolor1!30, fill=mycolor1!30] plot[smooth cycle, tension=0.5] coordinates
{(-0.3, -0.3) (-0.3, 1.3) (0.3, 1.3) (0.3, 0.3) (1.3, 0.3) (1.3, -0.3)};
\draw[draw=mycolor1!30, fill=mycolor1!30] plot[smooth cycle, tension=0.5] coordinates
{(1.7, -0.3) (1.7, 0.3) (2.7, 0.3) (2.7, 2.3) (3.3, 2.3) (3.3, -0.3)};
\draw (0, 0) -- (0, 1);
\draw (0, 0) -- (1, 0);
\draw (0, 2) -- (1, 2);
\draw (1, 1) -- (1, 2);
\draw (1, 1) -- (2, 1);
\draw (1, 2) -- (2, 2);
\draw (2, 1) -- (2, 2);
\draw (2, 0) -- (3, 0);
\draw (3, 0) -- (3, 1);
\draw (3, 1) -- (3, 2);
\draw[style=cut-edge] (0, 1) -- (0, 2);
\draw[style=cut-edge] (0, 1) -- (1, 1);
\draw[style=cut-edge] (1, 0) -- (1, 1);
\draw[style=cut-edge] (1, 0) -- (2, 0);
\draw[style=cut-edge] (2, 0) -- (2, 1);
\draw[style=cut-edge] (2, 1) -- (3, 1);
\draw[style=cut-edge] (2, 2) -- (3, 2);
\node[style=vertex] at (0, 0) {};
\node[style=vertex] at (1, 0) {};
\node[style=vertex] at (0, 1) {};
\node[style=vertex] at (0, 2) {};
\node[style=vertex] at (1, 1) {};
\node[style=vertex] at (1, 2) {};
\node[style=vertex] at (2, 1) {};
\node[style=vertex] at (2, 2) {};
\node[style=vertex] at (2, 0) {};
\node[style=vertex] at (3, 0) {};
\node[style=vertex] at (3, 1) {};
\node[style=vertex] at (3, 2) {};
\end{tikzpicture}
\caption{
Decomposition of a graph into three components (\textcolor{mycolor1}{green}).
The corresponding cut consists of the dashed edges straddling distinct components (\textcolor{red}{red}).
}
\label{fig:multicut}
\end{figure}
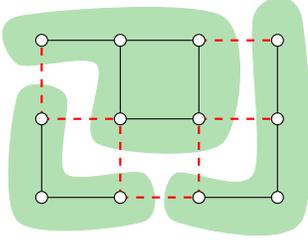

A \emph{decomposition (or clustering)} of a graph $G = (V,E)$ is a partition $\{V_1, \ldots, V_k\} $ of the node set such that $V_1 \cup \ldots, \cup V_k = V$ and $V_i \cap V_j = \varnothing$ $\forall i\neq j$.
The \emph{cut} $\delta(V_1,\ldots,V_k)$ induced by a decomposition is the subset of edges that straddle distinct clusters.
Such edges are said to be \emph{cut}.
See Figure~\ref{fig:multicut} for an illustration of a cut into three components.

The space of all multicuts is 
\begin{equation}
  \mathcal{M}_G = \left\{ \delta(V_1,\ldots,V_k) : 
\begin{array}{c}
  k \in \N \\
  V_1 \dot\cup \ldots \dot\cup V_k = V
\end{array} \right\}.
\end{equation} 

The associated minimum cost multicut problem is defined by an additional edge cost vector $c \in \R^{E}$.
For any edge $uv \in E$, negative costs $c_{uv} < 0$ favour the nodes $u$ and $v$ to be in distinct components.
Positive costs $c_{uv} > 0$ favour these nodes to lie in the same component.
The minimum cost multicut problem is
\begin{equation}
  \label{eq:multicut}
  \min_{y \in \mathcal{M}_G} \la c, y \ra,
\end{equation}
where $y_{uv}$ for edge $uv \in E$ is $1$ (resp. $0$) if $u$ and $v$ belong to distinct (resp. same) components. 

Below we detail the key components of our algorithm:
Starting from a graph where each node is a cluster, primal updates consist of edge contractions that iteratively merge clusters by join operations.
Dual updates optimize a Lagrange relaxation via message passing to obtain better edge costs and lower bound.
Primal and dual updates are interleaved to yield our primal-dual multicut algorithm.
We additionally detail how each operation can be done in a highly parallel manner.

\subsection{Primal: Parallel Edge Contraction}
The idea of edge contraction algorithms is to iteratively choose edges with large positive costs.
Such edges prefer their endpoints to be in the same component, hence they are contracted and end up in the same cluster.
Edge contraction is performed until no contraction candidates are found.
The special case of greedy additive edge contraction (GAEC)~\cite{keuper2015efficient} chooses in each iteration an edge with maximum edge weight for contraction and stops if each edge in the contracted graph has negative weight. 
The following Lemma describes the operation of edge contraction.
\begin{lemma}
\label{lemma:edge-contraction}
Let an undirected graph $G=(V,E,c)$ and a set of edges $S \subseteq E$ to contract be given. Also let $G'=(V', E', c')$ be the graph obtained after edge contraction.
\begin{enumerate}[label=(\alph*)]
\item
\label{item:edge-contraction-paths}
The corresponding surjective \emph{contraction mapping} $f : V \rightarrow V'$ mapping node set $V$ onto the contracted node set $V'$ is up to isomorphism uniquely defined by $f(u) = f(v) \iff \exists uv\textnormal{-path}(V,S)$. The contracted edge set is given by $E' = \{f(u)f(v): f(u) \neq f(v), uv \in E\}$.
\item
\label{item:edge-contraction-costs}
The edge weights for contracted edges are $c'_{ij} = \sum_{uv \in E: f(u) = i, f(v) = j} c_{uv}, \, \forall ij \in E'$.
\end{enumerate}
\end{lemma}

\begin{figure}
\centering
    \begin{tikzpicture}[scale=1.5,>=stealth]
\tikzstyle{join-edge}=[mycolor1,thick,dashed]
\tikzstyle{att-edge}=[black]
\tikzstyle{rep-edge}=[black]
\tikzstyle{vertex}=[circle, draw, fill=white, inner sep=1pt, minimum width=1ex]
\tikzset{every picture/.append style={baseline,scale=1.1}}
\tikzstyle{every node}=[font=\small]

\node[style=vertex](a) at (0,0) {$p$};
\node[style=vertex](c) at (-0.5,1) {$r$};
\node[style=vertex](d) at (0.5,1) {$s$};
\node[style=vertex](b) at (1,0) {$q$};
\node[style=vertex](e) at (1.5,1) {$t$};

\draw[style=att-edge] (a) -- node [above=-0.025cm] {$1$} (b);
\draw[style=rep-edge] (a) -- node [left=-0.025cm] {$-2$} (c);
\draw[style=rep-edge] (b) -- node [left=-0.025cm] {$-3$} (d);

\draw[style=att-edge] (b) -- node [left=-0.025cm] {$1$} (e);
\draw[style=join-edge] (c) -- node [above=-0.025cm] {$3$} (d);
\draw[style=join-edge] (d) -- node [above=-0.025cm] {$4$} (e);

\draw [out=60,in=120,looseness=0.75] (c.north) to node[above]{$1$} (e.north);

\node[draw=none](first_right_anchor) at (1.4,0.5) {};
\node[draw=none](second_left_anchor) at (2.3,0.5) {};

\draw[->,snake=zigzag, segment amplitude=.4mm, segment length=1mm, line after snake=1mm, gray] (first_right_anchor) -- (second_left_anchor);

\node[style=vertex](ap) at (2.5,0) {$p$};
\node[style=vertex](bp) at (3.5,0) {$q$};
\node[style=vertex](cp) at (3.0,1) {$r'$};

\draw[style=att-edge] (ap) -- node [below] {$1$} (bp);
\draw[style=rep-edge] (ap) -- node [left=-0.025cm] {$-2$} (cp);
\draw[style=rep-edge] (bp) -- node [right=-0.025cm] {$-3 + 1$} (cp);

\draw [out=120,in=60,looseness=10, densely dotted] (cp) to node[above]{$3+4+1$} (cp);

\end{tikzpicture}
\caption{
Contraction of a graph with contraction set $S=\{rs, st\}$ where vertices $r, s$ and $t$ are merged to form a cluster $r'$. The corresponding contraction mapping is $f(p) = p, f(q) = q, f(r)=f(s)=f(t)=r'$. Notice that edges $qs$ and $qt$ become parallel edges after contraction and their costs are added.
Also notice the presence of self-loop in the contracted graph with cost indicating intra-cluster similarity.
}
\label{fig:contraction}
\end{figure}
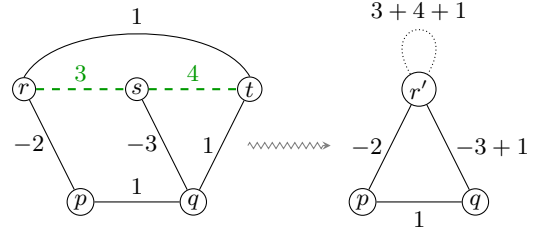
Lemma~\ref{lemma:edge-contraction}\ref{item:edge-contraction-paths} relates the contraction mapping $f$ with the set $S$ of edges to contract.
If two nodes have the same value in $f$ then there must be a path between them in graph $(V, S)$.
Moreover, the edges whose end points are not contracted are preserved in $E'$. Lemma~\ref{lemma:edge-contraction}\ref{item:edge-contraction-costs} provides the costs of contracted edges that are obtained by summing the costs of parallel edges. 
An illustration of the lemma is given in Figure~\ref{fig:contraction}.

In order to perform edge contraction fast we will use a linear algebraic representation that will allow to use highly parallel (sparse) matrix-matrix multiplication.
\begin{definition}[Adjacency Matrix]
Given a weighted graph $G=(V,E,c)$ its (symmetric) adjacency matrix $A \in \R^{V \times V}$ is defined by $A_{uv} = \begin{cases} c_{uv}, & uv \in E \\ 0,& \text{otherwise} \end{cases}$.
\end{definition}
We will perform edge contraction with the help of an edge contraction matrix defined as follows.
\begin{definition}[Edge Contraction Matrix]
Given a weighted graph $G=(V,E,c)$ and an edge set $S \subset E$ to contract,
let $f$ be the contraction mapping and $V'$ the contracted node set.
The edge contraction matrix $K_S \in \{0,1\}^{V \times V'}$ is defined as
$(K_{S})_{uu'} = \begin{cases}
1,& f(u) = u'\\
0,& \text{otherwise}
\end{cases}.$
\end{definition}
\begin{lemma}
\label{lemma:adjacency-matrix-contraction}
Given a weighted graph $G=(V,E,c)$, an edge set $S \subseteq E$ to contract and an associated edge contraction mapping $f$
\begin{enumerate}[label=(\alph*)]
\item
\label{item:adjacency-matrix-contraction}
the adjacency matrix of the contracted graph is equal to $K_S^{\top} A K_S - \diag(K_S^{\top} A K_S)$, where $\diag(\cdot)$ is the diagonal part of a matrix,
\item
\label{item:diagonal-entry-contraction}
it holds for the diagonal entry $(K_S^{\top} A K_S)_{u'u'} = \sum_{uv \in E: u' = f(u) = f(v)} c_{uv}$.
\end{enumerate}
\end{lemma}

Lemma~\ref{lemma:adjacency-matrix-contraction}\ref{item:adjacency-matrix-contraction} provides a way to compute the contracted graph in parallel by sparse matrix-matrix multiplication.
Lemma~\ref{lemma:adjacency-matrix-contraction}\ref{item:diagonal-entry-contraction} allows to efficiently judge whether the newly formed clusters decrease the multicut objective. Specifically if the diagonal contains all positive terms then the corresponding multicut objective will also decrease after contraction. 

A primal update iteration is given in Algorithm~\ref{alg:edge-contraction} that performs edge contraction as in Lemma~\ref{lemma:adjacency-matrix-contraction}\ref{item:adjacency-matrix-contraction}.

\begin{algorithm}
\DontPrintSemicolon
\caption{\texttt{Parallel-Edge-Contraction}}
\label{alg:edge-contraction}
\KwData{Graph $G=(V,E,c)$}
\KwResult{Contracted Graph $G' = (V',E',c')$, contraction mapping $f:V \rightarrow V'$}
Compute contraction set $S \subseteq E$\;
Compute adjacency matrix $A$ from $G$\;
Construct contraction mapping $f: V \rightarrow V'$\;
Construct contraction matrix $K_S$\;
$A' = K_S^{\top} A K_S - \diag(K_S^{\top} A K_S)$\;
Compute contracted graph $G'=(V',E',c')$ from $A'$\;
\end{algorithm}

\myparagraph{Finding contraction edge set $S$:} 
A vital step for ensuring a good primal update is selecting the edge set $S$ for contraction in Algorithm~\ref{alg:edge-contraction}. On one hand, we would like to choose edges in a conservative manner to avoid erroneous contractions. On the other hand, we need to contract as much edges as possible for efficiency. We propose two approaches allowing us to be at the sweet spot for both criterion as follows.
\begin{packed_description}
\item[Maximum matching:] Perform a fast maximum matching on the positive edges in using a GPU version of the Luby-Jones handshaking algorithm~\cite{cohen2012graph} and select the matched edges for contraction. 
\item[Maximum spanning forest without conflicts:] 
Compute a maximum spanning forest on the positive edges with a fast GPU version of Bor\r{u}vka's algorithm~\cite{wen2011gpu_gems} to find initial contraction candidates. Afterwards, iterate over all negative edges $ij$, find the unique path between $i$ and $j$ in the forest (if it exists) and remove the smallest positive edge.
We make use of GPU connected components~\cite{ecl_cc2018} to check for presence of these paths and to compute the final contraction mapping.
\end{packed_description}
Both of the above strategies ensure that the resulting join operation decreases the multicut objective.
We first find contraction edges via maximum matching. If not enough edges are found (\ie fewer than $0.1 \abs{V}$), we switch to the spanning forest based approach. Note that if we chose only one largest positive edge for contraction, Algorithm~\ref{alg:edge-contraction} specializes to GAEC~\cite{keuper2015efficient}. Since our algorithm depends upon many simultaneous edge contractions for efficiency, we do not use this strategy.

\subsection{Dual: Conflicted Cycles \& Message passing}
Solving a dual of multicut problem~\eqref{eq:multicut} can help in obtaining a lower bound on the objective value and also yields a reparametrization of the edge costs which can help in better primal updates.
Our dual algorithm works on the cycle relaxation for the multicut problem~\cite{chopra1993partition}.
We present for its solution massively parallel inequality separation routines to search for the most useful violated constraints and efficient dual block coordinate ascent procedure for optimizing the resulting relaxation.

\subsubsection{Cycle Inequalities \& Lagrange Relaxation}
Since the multicut problem is NP-hard~\cite{bansal2004correlation,demaine2006correlation}, we cannot hope to obtain a feasible polyhedral description of $\conv(\mathcal{M}_G)$.
A good relaxation for most practical problems is given in terms of cycle inequalities.
Given a cycle $C = \{e_1,\ldots,e_l\} \subseteq E$, a feasible multicut must either not contain any cut edge or should contain at least two cut edges. This constraint is expressed as
\begin{equation}
  \label{eq:cycle-inequality}
\forall C \in \text{cycles}(G):\   \forall e \in C: y_e \leq \sum_{e' \in C\backslash \{e\}} y_{e'}.
\end{equation}

Cycle inequalities together with the binary constraints $y_e \in \{0,1\}$ actually define $\mathcal{M}_G$~\cite{chopra1993partition}.
In other words, when relaxing $y_e \in [0,1]$ we obtain a linear program relaxation to $\conv(\mathcal{M}_G)$ with all integral points being valid multicuts.

While cycle inequalities~\eqref{eq:cycle-inequality} give us a polyhedral relaxation of the multicut problem~\eqref{eq:multicut}, our algorithm will operate on a Lagrangean decomposition that was proposed in~\cite{swoboda2017message}.
It consists of two types of subproblems joined together via Lagrange variables:
(i)~edge subproblems for each edge $e \in E$ and
(ii)~triangle subproblems (\ie cycles of length $3$) for a subset of triangles $T \subset \begin{pmatrix} E \\ 3 \end{pmatrix}$.
Triangulation of cycles of length more than three is done to get triangles defining the same polyhedral relaxation as the one with all possible cycle inequalities~\eqref{eq:cycle-inequality} without loss of generality~\cite{chopra1993partition}.
We define the set of feasible multicuts on triangle graphs as
\begin{equation}
    \M_{T} = \left\{ (0,0,0), (1,1,0), (1,0,1), (0,1,1), (1,1,1)\right\},
\end{equation}
which is a special case of~\eqref{eq:cycle-inequality} representing that either all edges are cut/joined or exactly two edges are cut. 
Given a set of edge and triangle subproblems our Lagrange decomposition is
\begin{equation}
\label{eq:dual-multicut}
\max_{\lambda} 
\underbrace{    
    \sum_{uv \in E} \min_{y \in \{0,1\}}c^{\lambda}_{uv} \cdot y
    + \sum_{t \in T} \min_{y \in \M_{T}} \la c_t^{\lambda}, y \ra
    }_{ =: \text{LB}(\lambda)}
\end{equation}
where the \emph{reparametrized} edge costs $c^{\lambda}_{uv} \in \mathbb{R}$ and triangle costs $c_t^{\lambda} \in \mathbb{R}^3$ for triangle $t = \{ij, jk, ki\} \in T$ are 
\begin{subequations}
\label{eq:reparametrization}
\begin{align}
   c_{uv}^{\lambda} &= c_{uv} + \sum_{t \in T: {uv} \in t} \lambda_{t,{uv}} \label{eq:reparametrized_edge_costs} \\ 
    c_t^{\lambda} &= -(\lambda_{t,ij}, \lambda_{t,jk}, \lambda_{t,ki}) 
\end{align}
\end{subequations}
$\text{LB}(\lambda)$ in~\eqref{eq:dual-multicut} is a lower bound on the cost of the optimum multicut for any $\lambda$.
The optimum objective value of~\eqref{eq:dual-multicut} equals that of the polyhedral relaxation~\cite{swoboda2017dual}.

\subsubsection{Cycle Inequality Separation}
\label{sec:cycle_sep}
For the dual problem~\eqref{eq:dual-multicut} one would need to enumerate all possible cycle inequalities~\eqref{eq:cycle-inequality}.
However, as mentioned in~\cite{lange2018partial} we can restrict only to conflicted cycles of $G$ for efficiency without loosening the relaxation.
A cycle is called a conflicted cycle if it contains exactly one repulsive edge.

\begin{definition}[Conflicted cycles]
Let the set of attractive edges in E be $E^+ =\{c_e > 0: \forall e \in E\}$ and repulsive edges $E^- = \{c_e < 0: \forall e \in E\}$. Then conflicted cycles of G is the set $\{C \in \text{cycles}(G): \lvert C \cap E^- \rvert\ = 1\}$.
\end{definition}
\begin{lemma}
\label{lemma:conf_cycle_search}
The search for conflicted cycles can be performed in parallel for each $ij \in E^-$ by finding shortest path w.r.t.\ hop distance between $i$ and $j$ in the graph $G=(V,E^+)$ making good use of parallelization capabilities of GPUs.
\end{lemma}
\subsubsection{Dual Block Coordinate Ascent (DBCA)}
DBCA (a.k.a.\ message passing) was studied in~\cite{swoboda2017message} for multicut.
However, the resulting message passing schemes are not easily parallelizable.
The underlying reason for the inherent sequential nature of these schemes is that the effectiveness of the proposed message passing operations depend on the previous ones being executed.
We propose a message passing scheme for multicut that is invariant to the message passing schedule, hence allowing parallel computation.

As in~\cite{swoboda2017message}, our scheme iteratively improves the lower bound~\eqref{eq:dual-multicut} by message passing between edges and triangles. 

For each message passing operation we need to compute min-marginals, \ie\ the difference of optimal costs on subproblems obtained by fixing a specified variable to $1$ and $0$.
For edge costs the min-marginal is just the reparametrized edge cost.
For triangle subproblems it is given as follows

\begin{definition}[Marginalization for triangle subproblems]
Let $t \in T$ be a triangle containing an edge $e$.
\begin{equation}
    m_{t \rightarrow e}(c_t^{\lambda}) = 
     \min_{\substack{y_e=1 \\ y \in \M_{T}}} \la c_t^{\lambda}, y\ra
    - \min_{\substack{y_e=0 \\ y \in \M_{T}}} \la c_t^{\lambda}, y\ra
\end{equation}
is called \emph{min-marginal} for triangle $t$ and edge $e$.

\end{definition}

The message passing algorithm iteratively sets min-marginal to zero first for edge subproblems and then for triangles described in Algorithm~\ref{alg:parallel-message-passing}. By sending back and forth messages the subproblems communicate their local optima and ultimately the min-marginals converge towards agreement (\ie their corresponding edge labels $y$ are consistent).
In~\cite{swoboda2017dual} it was shown that each such operation is non-decreasing in the dual objective value, yielding an overall monotonic convergence.
Message are passed from edges to triangles in lines~\ref{alg:edge-to-triangle-start}-\ref{alg:edge-to-triangle-end}.
After this step the reparametrized edge costs $c_e^\lambda$ become zero. 
We perform multiple triangle to edge message passing updates (line~\ref{alg:triangle-to-edge-start}-\ref{alg:triangle-to-edge-end}) similar to the way it was done in~\cite{tourani2018mplp++} that distribute messages uniformly among all triangles which contain that edge.
After this operation min-marginals for $c^{\lambda}_t$ become zero.

\begin{algorithm}
\DontPrintSemicolon
\caption{\texttt{Parallel-Message-Passing}}
\label{alg:parallel-message-passing}
\KwData{Graph $G=(V,E,c)$, triangles $T$, Lagrange multipliers $\lambda$.}
\KwResult{Updated Lagrange multipliers $\lambda$}
\tcp{Messages from edges to triangles}
\For{$e \in E$ in parallel} {
$\alpha = c^{\lambda}_e$ \;
\label{alg:edge-to-triangle-start}
\For{$t \in T: e \in t$} {
$\lambda_{t, e} -= \frac{\alpha}{\abs{t \in T: e \in t}}$ \;
}
\label{alg:edge-to-triangle-end}
}
\tcp{Messages from triangles to edges}
\For{$t = \{ij, jk, ki\} \in T$ in parallel} {
$\lambda_{t, ij} += \frac{1}{3} m_{t \rightarrow ij}(c^{\lambda}_t) $\;
\label{alg:triangle-to-edge-start}
$\lambda_{t, ik} += \frac{1}{2} m_{t \rightarrow ik}(c^{\lambda}_t) $\;
$\lambda_{t, jk} += m_{t \rightarrow jk}(c^{\lambda}_t) $ \;

$\lambda_{t, ij} += \frac{1}{2} m_{t \rightarrow ij}(c^{\lambda}_t) $\; 
$\lambda_{t, ik} += m_{t \rightarrow ik}(c^{\lambda}_t) $ \;
$\lambda_{t, ij} += m_{t \rightarrow ij}(c^{\lambda}_t) $ \; 
\label{alg:triangle-to-edge-end}
}
\end{algorithm}

\myparagraph{Convergence of Message Passing.}
Algorithm~\ref{alg:parallel-message-passing} converges towards fixed points, similar to other DBCA schemes for graphical models~\cite{tourani2018mplp++,werner2007linear,kolmogorov2005convergent,kolmogorov2014new}. 
These fixed points are characterized with the help of arc consistency and need not coincide with the optimal dual solution, but are typically close to them.
Below, we characterize these fixed points.

\begin{definition}[Locally Optimal Solutions]
Define the locally optimal solutions for edge $e \in E$ as
\begin{equation}
    \overline{c^{\lambda}_e} := \{ x \in \{0,1\} : x \cdot c^{\lambda}_e = \min(0,c^{\lambda}_e) \}
\end{equation}
and similarly for triangle $t \in T$ as 
\begin{multline}
    \overline{c^{\lambda}_t} := \{ x \in \M_{T} :  \la c^{\lambda}_t, x\ra = \min_{x' \in \M_{T}} \la c^{\lambda}_t, x' \ra \}\,
\end{multline}
Define the projection of triangle solutions onto one of its edges as 
\begin{equation}
    \Pi_e(\overline{c^{\lambda}_t}) := \{x \in \{0,1\} : \exists x' \in \overline{c^{\lambda}_t} \text{ s.t.\ } x'_e = x \}\,
\end{equation}
\end{definition}
\begin{definition}[Arc-Consistency]
Lagrange multipliers $\lambda$ are arc-consistent if $\Pi_e(\overline{c^{\lambda}_t}) = \overline{c^{\lambda}_e}$ for all $t \in T$ and $e \subset t$.
\end{definition}
However, note that arc-consistency is not necessary for dual optimality.
A necessary criterion is edge-triangle agreement.
\begin{definition}[Edge-Triangle Agreement]
Lagrange multipliers are in edge-triangle agreement if there exist non-empty subsets $\xi_e \subseteq \overline{c^{\lambda}_e}$ for all $e \in E$ and $\xi_t \subseteq \overline{c^{\lambda}_t}$ for all $t \in T$ such that $\xi$ is arc-consistent, i.e.\ $\xi_e = \Pi_e(\xi_t)$ for all $t \in T$ and $e \subset t$.
\end{definition}
In words, edge-triangle agreement signifies that there exists a subset (also called kernel in~\cite{werner2007linear}) of locally optimal solutions that are arc-consistent.
\begin{theorem}
\label{thm:edge-triangle-agreement-convergence}
Algorithm~\ref{alg:parallel-message-passing} converges to edge-triangle agreement.
\end{theorem}

\subsection{Primal-Dual Updates}
\begin{figure}
\centering
    \begin{tikzpicture}[scale=1.5,>=stealth]
\def\circledarrow#1#2#3{ 
\draw[#1,->] (#2) +(80:#3) arc(80:-260:#3);
}
\tikzstyle{att-edge}=[scattercolor3!60, line width=0.3mm]
\tikzstyle{rep-edge}=[scattercolor5!60, line width=0.3mm]
\tikzstyle{vertex}=[circle, draw, fill=white, inner sep=1pt, minimum width=1ex]
\tikzset{every picture/.append style={baseline,scale=1.1}}
\tikzstyle{every node}=[font=\small]

\node[style=vertex](a) at (0,0) {$a$};
\node[style=vertex](b) at (0.8,0) {$b$};
\node[style=vertex](c) at (1.3,0.7) {$c$};
\node[style=vertex](d) at (0.6,1.1) {$d$};
\node[style=vertex](e) at (-0.2,1) {$e$};

\draw[style=att-edge, line width=1mm] (a) -- (b);
\draw[style=att-edge, line width=1mm] (b) -- (c);
\draw[style=rep-edge, line width=1.5mm] (c) -- (d);
\draw[style=att-edge, line width=1mm] (d) -- (e);
\draw[style=rep-edge, line width=1.5mm] (e) -- (a);
\draw[style=att-edge, line width=1mm] (a) -- (d);
\draw[style=att-edge, line width=1mm] (b) -- (d);

\node(t1) at (0.15, 0.7) {};
\circledarrow{gray}{t1}{0.12cm};

\node(t2) at (0.5, 0.45) {};
\circledarrow{gray}{t2}{0.12cm};

\node(t3) at (0.95, 0.6) {};
\circledarrow{gray}{t3}{0.12cm};

\node[draw=none](first_right_anchor) at (1.35,0.5) {};
\node[draw=none](second_left_anchor) at (1.75,0.5) {};

\draw[->,snake=zigzag, segment amplitude=.4mm, segment length=1mm, line after snake=1mm, gray]  (first_right_anchor) -- node [below, label={[below,align=center,font=\tiny]Message\\ passing}] {} (second_left_anchor);

\def\firstskip{2.0}
\node[style=vertex](a1) at (0 + \firstskip,0) {$a$};
\node[style=vertex](b1) at (0.8 + \firstskip,0) {$b$};
\node[style=vertex](c1) at (1.3 + \firstskip,0.7) {$c$};
\node[style=vertex](d1) at (0.6 + \firstskip,1.1) {$d$};
\node[style=vertex](e1) at (-0.2 + \firstskip,1) {$e$};

\draw[style=att-edge, line width=0.5mm] (a1) -- (b1);
\draw[style=att-edge, line width=0.5mm] (b1) -- (c1);
\draw[style=rep-edge, line width=0.5mm] (c1) -- (d1);
\draw[style=att-edge, line width=0.5mm] (d1) -- (e1);
\draw[style=rep-edge, line width=0.5mm] (e1) -- (a1);
\draw[style=rep-edge, line width=0.5mm] (a1) -- (d1);
\draw[style=rep-edge, line width=0.5mm] (b1) -- (d1);

\node[draw=none](first_right_anchor1) at (1.35 + \firstskip,0.5) {};
\node[draw=none](second_left_anchor1) at (1.75 + \firstskip,0.5) {};

\draw[->,snake=zigzag, segment amplitude=.4mm, segment length=1mm, line after snake=1mm, gray] (first_right_anchor1) -- node [below, label={[below,align=center,font=\tiny]Edge\\ contraction}] {}  (second_left_anchor1);

\def\secondskip{3.8}
\node[style=vertex](abc) at (0.4 + \secondskip,0) {$abc$};
\node[style=vertex](de) at (0.4 + \secondskip, 1.1) {$de$};

\draw[style=rep-edge, line width=2mm] (abc) -- (de);







\end{tikzpicture}
\caption{Example iteration of our primal-dual multicut solver on a graph with \textcolor{scattercolor5}{repulsive} and \textcolor{scattercolor3}{attractive} edges. Width of the edges indicate abs. cost. First we detect conflicted cycles and triangulate to get triangles (indicated by $\circlearrowright$). Next, dual update reparametrizes edge costs which resolves the conflicted cycles. Lastly a primal update is done by contracting attractive edges.}
\label{fig:primal-dual}
\end{figure}
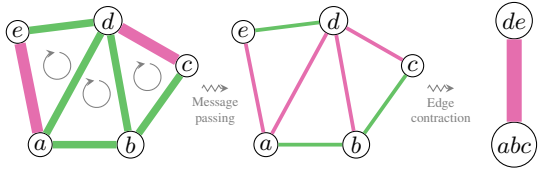

While the two building blocks of our multicut solver \ie edge contraction and cycle separation with message passing can be used in isolation to compute a primal solution and lower bound, we propose an interleaved primal-dual solver in Algorithm~\ref{alg:primal-dual-parallel-multicut}.

\SetInd{0.25em}{0.75em} 
\begin{algorithm}
\caption{\texttt{Primal-Dual Multicut}}
\label{alg:primal-dual-parallel-multicut}

\DontPrintSemicolon
\KwData{Graph $G=(V,E,c)$}
\KwResult{Contraction mapping $f : V \rightarrow V'$}
\tcp{Init.\ each node as a separate cluster}
$f = V \rightarrow V$, $f(v) = v$ \ \ $\forall v \in V$\;
\While{$G$ has positive edges without conflicts}{
\tcp{Find conflicted cycles (Lemma.~\ref{lemma:conf_cycle_search}) }
$T = \texttt{Cycle-Separation}(G)$\;
\For{iter= $1,\ldots,k$}{
$\lambda = \texttt{Parallel-Message-Pass.}(G, T)$\;
\tcp{Reparametrize edge costs}
$c_e = c_e^{\lambda}\quad \forall e \in E$ by ~\eqref{eq:reparametrized_edge_costs}\label{alg:pd_alg_reparam}\;
}
$G, f' =\texttt{Parallel-Edge-Contract.}(G)$\label{alg:pd_contract}\;
\tcp{Update contraction mapping}
$f(v) = f'(f(v))$ $\forall v \in V$\;
}
\end{algorithm}

In each iteration we separate cycles and perform message passing to get reparameterized edge costs. We use these reparametrized edge costs to perform parallel edge contraction. This interleaved process continues until no edge contraction candidate can be found.
Such scheme has the following benefits
\begin{packed_description}
\item[Better edge contraction costs:]
The reparametrization in line~\ref{alg:pd_alg_reparam} produces edge costs $c^{\lambda}$ that are more indicative of whether an edge is contracted or not in the final solution thus yielding better primal updates in line~\ref{alg:pd_contract}.
In case the relaxation~\eqref{eq:dual-multicut} is tight, the sign of $c^{\lambda}_e$ perfectly predicts whether an edge $e$ is separating two clusters or is inside one.
\item[Better cycle separation:]
For fast execution times we stop cycle separation for cycles greater than a given length ($5$ in our case).
Since cycle separation is performed again after edge contraction, this corresponds to finding longer cycles in the original graph. Such approach alleviates the need to perform a more exhaustive and time-consuming initial search. 
\end{packed_description}
Note that a valid lower bound can be obtained from Algorithm~\ref{alg:primal-dual-parallel-multicut} by recording~\eqref{eq:dual-multicut} after cycle separation and message passing on the original graph. 
\section{Experiments}
\label{sec:experiments}
We evaluate solvers on multicut problems for neuron segmentation for connectomics in the fruit-fly brain~\cite{pape2017solving} and unsupervised image segmentation on Cityscapes~\cite{cordts2016cityscapes}.
We use a single NVIDIA Volta V100 (16GB) GPU for our solvers unless otherwise stated and an AMD EPYC 7702 for CPU solvers. Our solvers are implemented using the CUDA~\cite{cuda} and Thrust~\cite{Thrust} GPU programming frameworks.

\myparagraph{Datasets}
We have chosen three datasets containing the largest multicut problem instances we are aware of. The instances are available in~\cite{swoboda2022sp_archive}. 
\begin{packed_description}
\item[\normalfont\textit{Connectomics-SP}:]
Contains neuron segmentation problems from the fruit-fly brain~\cite{pape2017solving}.
The raw data is taken from the CREMI-challenge~\cite{cremi} acquired by~\cite{zheng2018complete} and converted to multiple multicut instances by~\cite{pape2017solving}. For this conversion~\cite{pape2017solving} also reduced the problem size by creating super-pixels.
The majority of these instances are different crops of one global problem.
There are 3 small ($400000-600000$ edges), 3 medium ($4-5$ million edges) and 5 large ($28-650$ million edges) multicut instances.
For the largest problem we use NVIDIA RTX 8000 (48GB) GPU.
\item[\normalfont\textit{Connectomics-Raw}:]
We use the $3$ test volumes (sample A+, B+, C+) from the CREMI-challenge~\cite{cremi} segmenting directly on the pixel level without conversion to super-pixels. Conversion to multicut instances is carried out using~\cite{torch_EM}.
We report results on two types of instances: 
(i)~The three full problems where the underlying volumes have size $1250 \times 1250 \times 125$ with around $700$ million edges and
(ii)~six cropped problems created by halving each volume and creating the corresponding multicut instances each containing almost $340$ million edges.
For all these instances we use NVIDIA RTX 8000 (48GB) GPU. 
\item[\normalfont\textit{Cityscapes}:]
Unsupervised image segmentation on $59$ high resolution images ($2048 \times 1024$) taken from the validation set~\cite{cordts2016cityscapes}.
Conversion to multicut instances is done by computing the edge affinities produced by~\cite{abbas2021combinatorial} on a grid graph with $4$-connectivity and additional coarsely sampled longer range edges.
Each instance contains approximately $2$ million nodes and $9$ million edges. 
\end{packed_description}

\myparagraph{Algorithms}
As baseline methods we have chosen, to our knowledge, the fastest primal heuristics from the literature.
\begin{packed_description}
\item[\normalfont\texttt{GAEC}~\cite{keuper2015efficient}:] The greedy additive edge contraction corresponds to 
Algorithm~\ref{alg:parallel-message-passing} with choosing a single highest edge to contract.
We use our own CPU implementation that is around $1.5$ times faster than the one provided by the authors.
\item[\normalfont\texttt{KLj}~\cite{keuper2015efficient}:] The  Kernighan\&Lin with joins algorithm performs local move operations which can improve the objective. To avoid large runtimes the output of \texttt{GAEC} is used for initialization. 
\item[\normalfont\texttt{GEF}~\cite{levinkov2017comparative}:] The greedy edge fixation algorithm is similar to \texttt{GAEC} but additionally visits negative valued (repulsive) edges and adds non-link constraints between their endpoints.
\item[\normalfont\texttt{BEC}~\cite{kardoost2018solving}:] Balanced edge contraction, a variant of \texttt{GAEC} which chooses edges to contract based on their cost normalized by the size of the two endpoints. 
\item[\normalfont\texttt{ICP}~\cite{lange2018partial}:] The iterated cycle packing algorithm searches for cycles and greedily solves a packing problem that approximately solves the multicut dual~\eqref{eq:dual-multicut}.
\item[\normalfont\texttt{P}:] Our purely primal Algorithm~\ref{alg:edge-contraction} using the maximum matching and spanning forest based edge contraction strategy.
\item[\normalfont\texttt{PD}:]
Our primal-dual Algorithm~\ref{alg:primal-dual-parallel-multicut} which additionally makes use of the dual information. We find conflicted cycles up to length $5$ on original graph and up to a length of $3$ for later iterations on contracted graphs. 
\item[\normalfont\texttt{PD+}:]
Variant of \texttt{PD} which always considers conflicted cycles up to a length $5$ for reparametrization which can lead to even better primal solutions although with higher runtime.
\item[\normalfont\texttt{D}:] Our dual cycle separation algorithm followed by message passing on the original graph via Algorithm~\ref{alg:parallel-message-passing} producing lower bounds. 
\end{packed_description}

\begin{table*}
    \centering
    \resizebox{\textwidth}{!}{\begin{tabular}{l rr rr rr rr rr rr}
    \toprule
    & \multicolumn{6}{c}{\normalfont\textit{Connectomics-SP}} & \multicolumn{4}{c}{\normalfont\textit{Connectomics-Raw}} & \multicolumn{2}{c}{\normalfont\textit{Cityscapes}} \\
    \cmidrule(lr){2-7} \cmidrule(lr){8-11} \cmidrule(lr){12-13}
    & \multicolumn{2}{c}{Small (3)} & \multicolumn{2}{c}{Med. (3)} & \multicolumn{2}{c}{Large (5)} & \multicolumn{2}{c}{Crops (6)} & \multicolumn{2}{c}{Full (3)} & \multicolumn{2}{c}{(59)} \\
    \cmidrule(lr){2-3} \cmidrule(lr){4-5} \cmidrule(lr){6-7} \cmidrule(lr){8-9} \cmidrule(lr){10-11} \cmidrule(lr){12-13}
    Method & C$(\times 10^5)$ & t(s) & C$(\times 10^5)$ & t(s) & C$(\times 10^5)$ & t(s) & C$(\times 10^8)$ & t(s) & C$(\times 10^8)$ & t(s) & C$(\times 10^6)$ & t(s)\\ 
    \midrule
    \multicolumn{13}{c}{Primal} \\
    \midrule

    \texttt{KLj}~\cite{keuper2015efficient} & $ -\textbf{1.794} $ & $3.8$ & $ -\textbf{9.225} $ & $125$ & $\dagger$ & $\dagger$ & $\dagger$ & $\dagger$ & $\dagger$ & $\dagger$ & $-1.858$ & $5e4$ \\ 
    \texttt{GAEC}~\cite{keuper2015efficient} & $ -1.794 $ & $0.4$ & $ -9.224 $ & $4.7$ & $ -\textbf{1.512} $ & $280$  & $-1.464$
& $570$ & $-2.963$ & $1140$ & $-1.826$ & $13$ \\
    \texttt{GEF}~\cite{levinkov2017comparative} & $-1.793$ & $0.7$ & $-9.223$ & $9.0$ & $-1.511$ & $699$  & $-1.458
$ & $582$ & $-2.949$ & $1762$ & $-1.743$ & $14$ \\
    \texttt{BEC}~\cite{kardoost2018solving} & $-1.787$ & $0.5$ & $-9.199$ & $5.6$ & $-1.507$ & $309$  & $-1.402
$ & $1688$ & $-2.838$ & $4150$ &  $-1.613$ & $36$ \\ 
    \texttt{P} & $-1.780$ & $\textbf{0.1}$ & $-9.173$ & $\textbf{0.6}$ & $-1.505$ & $\textbf{6}$  & $-1.430
$ & $\textbf{9}$ & $-2.895
$ & $\textbf{19}$ &  $-1.711$ & $\textbf{0.4}$ \\
    \texttt{PD} & $-1.791$ & $0.2$ & $-9.217$ & $1.0$ & $-1.509$ & $13$  & $-1.477
$ & $24$ & $-2.981$
 & $32$ &  $-1.846$ & $1$ \\
    \texttt{PD+} & $-1.791$ & $0.3$ & $-9.219$ & $1.4$ & $-1.509$ & $20$  & $-\textbf{1.480}
$ & $115$ & $-\textbf{2.995}$ & $224$ &  $-\textbf{1.862}$ & $2.2$ \\

    \midrule
    \multicolumn{13}{c}{Dual} \\
    \midrule
    \texttt{ICP}~\cite{lange2018partial} & $-1.798$ & $0.8$ & $-9.246$ & $11.3$ & $-1.518$ & $1235$ & $-1.507
$& $513$ & $-\textbf{3.053}$ & $\textbf{1091}$ & $-1.930$   &  $41.1$ \\
    \texttt{D} & $-\textbf{1.797}$ & $\textbf{0.2}$ & $-\textbf{9.241}$ & $\textbf{0.8}$ & $-\textbf{1.517}$ & $\textbf{13}$ & $-\textbf{1.499}
$ & $\textbf{34}$ & * & * & $-\textbf{1.928}$ &  $\textbf{1.3}$ \\
    \bottomrule
    \end{tabular}}
    \caption{Comparison of results on all datasets. (C: cost, t(s): time in seconds,
    $\dagger$: timed out, *: out of GPU memory). We report average primal and dual costs and runtime over instances within each category. In terms of primal solutions our primal-dual solvers (\texttt{PD}, \texttt{PD+}) achieve objectives close to or better than sequential solvers while being substantially faster especially on larger instances. Moreover, our parallel message passing approach (\texttt{D}) gives better lower bounds than \texttt{ICP} with up to two orders of magnitude reduction in runtime.}
    \label{tab:fruit-fly-cityscapes-results}
\end{table*}

\begin{figure}
\centering
    \input{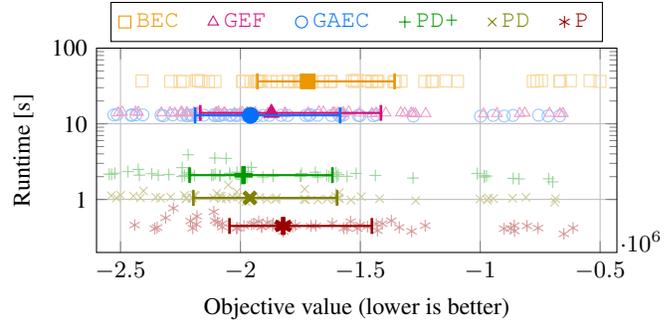}
\caption{Comparison of primal solutions from \normalfont\textit{Cityscapes} dataset. 
Our purely primal algorithm (\texttt{P}) is $30\times$ faster than \texttt{GAEC}~\cite{keuper2015efficient} and \texttt{GEF}~\cite{levinkov2017comparative}, although with worse objective values. Incorporating dual information enables our solvers (\texttt{PD}, \texttt{PD+}) to even surpass the sequential solvers in objective while being faster by an order of magnitude. Error bars mark the 0.25, 0.75-quantile. (\texttt{KLj} not shown due to high runtime). }
\label{fig:cityscapes_primal_scatter}
\end{figure}

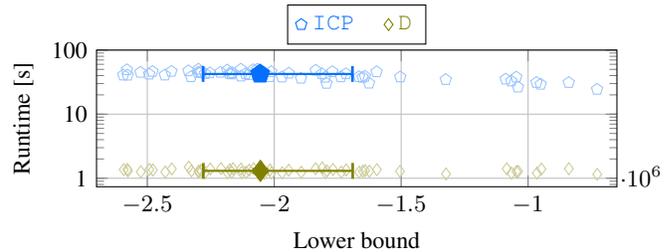
\begin{figure}
\centering
    \begin{tikzpicture}[font=\small]
\begin{axis}[
        width=8.5cm,
        height=3.4cm,
        yticklabels={, 1, 10, 100},
    	legend style={at={(0.5,1.04)},
	    anchor=south,legend columns=-1},
        xmajorgrids,
        ymajorgrids,
        xlabel={Lower bound},
        xmin=-2700000, xmax=-650000,
        ylabel={Runtime [s]},
        ylabel near ticks,
        xlabel near ticks,
        ymode=log,
        ymin=0, ymax=100,
        mark size=0.5ex,
        clip marker paths=true,
        every x tick scale label/.style={at={(1,0)},xshift=1pt,anchor=south west,inner sep=0pt}
    ]
    
    \addlegendimage{only marks, mark = pentagon, mark options={color=scattercolor6}}
    \addlegendentry{\textcolor{scattercolor6}{\texttt{ICP}\quad}}

    \addlegendimage{only marks, mark = diamond, mark options={color=scattercolor4}}
    \addlegendentry{\textcolor{scattercolor4}{\texttt{D}}}

    \addplot[
        only marks, mark=pentagon, color=scattercolor6!40
    ] table [
        col sep=comma, 
        skip first n=1,
        x index=1, y index=2,
        y expr=\thisrowno{2}*(1/1000),
    ] {Figures/scatterplots/cityscapes/icp_cityscapes.csv};
    
    \addplot[
        only marks, mark=diamond, color=scattercolor4!40
    ] table [
        col sep=comma, 
        skip first n=1,
        x index=2, y index=3,
        y expr=\thisrowno{3}*(1/1000),
    ] {Figures/scatterplots/cityscapes/cityscapes_5_10_5_only_lb.csv};


    \addplot[
        forget plot, only marks, mark=pentagon*, mark options={scale=1.3, ultra thick}, color=scattercolor6,
        error bars/.cd, x dir=plus, x explicit, y dir=plus, y explicit,
    ] table [
        col sep=comma, 
        skip first n=1,
        x index=0, y index=3,
        y expr=\thisrowno{3}*(1/1000),
        x error expr=\thisrowno{2}-\thisrowno{0}, 
    ] {Figures/scatterplots/cityscapes/icp_cityscapes_mean.csv};

    \addplot[
        forget plot, only marks, mark=pentagon*, mark options={scale=1.3, ultra thick}, color=scattercolor6,
        error bars/.cd, x dir=minus, x explicit, y dir=minus, y explicit
    ] table [
        col sep=comma, 
        skip first n=1,
        x index=0, y index=3,
        y expr=\thisrowno{3}*(1/1000),
        x error expr=\thisrowno{0}-\thisrowno{1}, 
    ] {Figures/scatterplots/cityscapes/icp_cityscapes_mean.csv};
    
    \addplot[
        forget plot, only marks, mark=diamond*, mark options={scale=1.3, ultra thick}, color=scattercolor4,
        error bars/.cd, x dir=plus, x explicit, y dir=plus, y explicit,
    ] table [
        col sep=comma, 
        skip first n=1,
        x index=3, y index=6,
        y expr=\thisrowno{6}*(1/1000),
        x error expr=(\thisrowno{5}-\thisrowno{3}),
        y error expr=(\thisrowno{8}-\thisrowno{6})*(1/1000),
    ] {Figures/scatterplots/cityscapes/cityscapes_5_10_5_only_lb_mean.csv};

    \addplot[
        forget plot, only marks, mark=diamond*, mark options={scale=1.3, ultra thick}, color=scattercolor4,
        error bars/.cd, x dir=minus, x explicit, y dir=minus, y explicit
    ] table [
        col sep=comma, 
        skip first n=1,
        x index=3, y index=6,
        y expr=\thisrowno{6}*(1/1000),
        x error expr=\thisrowno{3}-\thisrowno{4}, 
        y error expr=(\thisrowno{6}-\thisrowno{7})*(1/1000)
    ] {Figures/scatterplots/cityscapes/cityscapes_5_10_5_only_lb_mean.csv};
    

    
\end{axis}
\end{tikzpicture}
\caption{Comparison of lower bounds from \normalfont\textit{Cityscapes} dataset. Our parallel message passing scheme (\texttt{D}) is more than an order of magnitude faster than \texttt{ICP}~\cite{lange2018partial} and gives slightly better lower bounds. Error bars mark the 0.25, 0.75-quantile. }
\label{fig:cityscapes_dual_scatter}
\end{figure}

\begin{figure}
\centering
    \begin{tikzpicture}[font=\small]
\begin{axis}[
        width=8.5cm,
        height=3.25cm,
        xmajorgrids,
        ymajorgrids,
        xlabel={Instance size $\lvert E \rvert$},
        ylabel={Runtime [s]},
        ylabel near ticks,
        xlabel near ticks,
        mark size=0.5ex,
        legend pos=north east,
        clip marker paths=true,
        xmin=100000, xmax=550000000,
        scaled x ticks=base 10:-9,
        xtick scale label code/.code={},
        xticklabel={\pgfmathprintnumber{\tick}$B$},
        ytick={0, 250, 500, 750, 1000},
    	legend style={at={(0.5,1.04)},
	    anchor=south,legend columns=-1},
        every x tick scale label/.style={at={(1,0)},xshift=1pt,anchor=south west,inner sep=0pt}
    ]
    
    
    \addlegendimage{only marks, mark = halfcircle*, mark options={color=scattercolor6}}
    \addlegendentry{\textcolor{scattercolor6}{\texttt{GAEC}\quad}}

    \addlegendimage{only marks, mark = halfsquare*, mark options={color=scattercolor5}}
    \addlegendentry{\textcolor{scattercolor5}{\texttt{PD}\quad}}
        
    \addplot[mark=halfcircle*, color = scattercolor6, forget plot] table [
        col sep=comma, 
        skip first n=1,
        x index=3, y index=1,
        y expr=\thisrowno{1}*(1/1000),
    ] {Figures/scatterplots/cremi/gaec_cremi.csv};

    \addplot[mark=halfsquare*, color = scattercolor5, forget plot] table [
        col sep=comma, 
        skip first n=1,
        x index=5, y index=3,
        y expr=\thisrowno{3}*(1/1000)
    ] {Figures/scatterplots/cremi/CREMI_mst_thresh_new_param_5_10_3_5_1.csv};

    \node[draw = none] at (axis cs:169910665, 140) {\textcolor{gray}{\tiny $12s$}};

    \node[draw = none] at (axis cs:342265550, 150) {\textcolor{gray}{\tiny $22s$}};
    
    \node[draw = none] at (axis cs:520180270, 160) {\textcolor{gray}{\tiny $36s$}};

\end{axis}
\end{tikzpicture}
    \caption{Runtime scaling comparison computed on different crops of CREMI test data showing that RAMA scales very well as compared to \texttt{GAEC}~\cite{keuper2015efficient} w.r.t.\ increasing problem sizes}
\label{fig:cremi_scaling}
\end{figure}
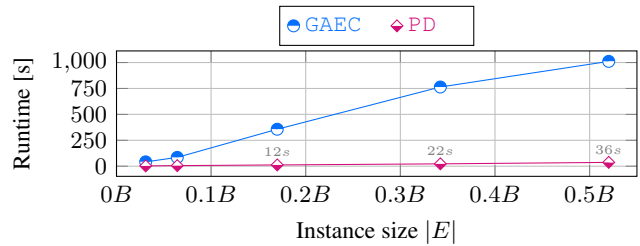
\pagebreak 

\myparagraph{Discussion}
Results on all datasets are given in Table~\ref{tab:fruit-fly-cityscapes-results}. 
On the \normalfont\textit{Connectomics-SP} dataset we attain primal objectives very close to those produced by \texttt{GAEC}~\cite{keuper2015efficient} but faster by more than an order of magnitude on large instances. 

For the \normalfont\textit{Cityscapes} and \normalfont\textit{Connectomics-Raw} datasets we achieve even better primal solutions than sequential algorithms by incorporating dual information while also being substantially faster. Our best solver (\texttt{PD+}) is more than $10^4$ times faster than \texttt{KLj}~\cite{keuper2015efficient} and produces better solutions.
Distributions of runtimes and primal resp.\ dual objectives for all instances of \normalfont\textit{Cityscapes} are shown in Figures~\ref{fig:cityscapes_primal_scatter} and \ref{fig:cityscapes_dual_scatter}.
We compare the scaling behaviour of our solver w.r.t increasing instance sizes in Figure~\ref{fig:cremi_scaling} showing that RAMA scales much more efficiently than \texttt{GAEC}. An example visual comparison of results is given in Figure~\ref{fig:cityscapes_results_comparison} in Appendix.

Lastly, our dual algorithm (\texttt{D}) produces speedups of up to two orders of magnitude and better lower bounds compared to the serial \texttt{ICP}~\cite{lange2018partial}, except on the full instances of \normalfont\textit{Connectomics-Raw} where we run out of GPU memory. 

\myparagraph{Runtime breakdown}
Runtime breakdown of our \texttt{PD} algorithm is given in Table~\ref{tab:time_breakdown}. Most of the time is spent in finding conflicted cycles which we found to be challenging to implement on GPU while keeping runtime and memory consumption low. Future improvements offer a potential for even better results and speedups by finding longer cycles more efficiently. 
\begin{table}[h!]
\begin{tabular}{c c c c}
\toprule
Finding $S$ & Contract. & Conf. cycles & Message passing \\
\cmidrule(lr){1-1} \cmidrule(lr){2-2} \cmidrule(lr){3-3} \cmidrule(lr){4-4}
$30\%$ & $7\%$ & $43\%$ & $20\%$ \\ 
\bottomrule
\end{tabular}
\caption{Runtime breakdown for \texttt{PD} algorithm on \textit{Cityscapes}}
\label{tab:time_breakdown}
\end{table}

\section{Conclusion}
\label{sec:conclusion}
We have demonstrated that multicut, an important combinatorial optimization problem for machine learning and computer vision, can be effectively parallelized on GPU. Our approach produces better solutions than state of the art efficient heuristics on grid graphs and comparable ones on super-pixel graphs while being faster by one to two orders-of-magnitude. We believe that performance gap on super-pixel graphs is due to a graph structure containing much more (and longer) conflicted cycles. Since our implementation can only find cycles of length up to 5, better implementations that can efficiently handle longer cycles might yield further improvements.

We estimate that the runtime gap will even widen in the future with the ever-increasing computing power of GPUs as compared to CPUs.
In contrast to CPU algorithms, where execution speed is the limiting factor, for our GPU algorithm, comparatively smaller amount of GPU-memory limits application to even larger instances.
We hope that our work will enable more compute intensive applications of multicut, where until now the slower serial CPU codepath has hindered its adoption. It might be possible to overcome GPU-memory limitations by multi-GPU implementations and/or decomposition methods.
\section*{Acknowledgments}
We would like to thank Shweta Mahajan and Jan-Hendrik Lange for insightful discussions and Constantin Pape, Adrian Wolny and Anna Kreshuk for their help and valuable suggestions regarding the experiments.
We would also like to thank all anonymous reviewers for their valuable feedback. 

{\small
\bibliographystyle{ieee_fullname}
\bibliography{references}
}

\section{Appendix}
\label{sec:appendix}
\subsection{Proof of Theorem~\ref{thm:edge-triangle-agreement-convergence}.}

The proofs are a condensation and adaptation of the corresponding proofs in~\cite{tourani2018mplp++,kolmogorov2005convergent,kolmogorov2014new}.
Changes are necessary since Algorithm~\ref{alg:parallel-message-passing} solves a different problem and uses different message passing updates and schedules than the algorithms from~\cite{tourani2018mplp++,kolmogorov2005convergent,kolmogorov2014new}.
As a shorthand we will use $c^{\lambda}_t (y)$ instead of writing $\la c^{\lambda}_t, y \ra$ for a solution $y$ of triangle subproblem $t \in T$.
\begin{definition}[$\epsilon$-optimal local solutions]
For $e \in E$ define 
\begin{equation}
\OO^{\epsilon}_e(\lambda) := \{x \in \{0,1\} : x \cdot c^{\lambda}_e \leq \min(0,c^{\lambda}_e) + \epsilon\}
\end{equation}
and for $t \in T$
\begin{equation}
\OO^{\epsilon}_t(\lambda) := \{x \in \M_{T} : c^{\lambda}_t(x) \leq \min_{x' \in \M_{T}} c^{\lambda}_t(x') + \epsilon\}
\end{equation} 
to be the $\epsilon$- optimal local solutions.
\end{definition}
Hence, $\OO^0_e(\lambda) = \overline{c^{\lambda}_e}$ for $e \in E$ and likewise $\OO^0_t(\lambda) = \overline{c^{\lambda}_t}$ for $t \in T$.
\begin{definition}[$\epsilon$-tolerance]
The minimal value $\epsilon(\lambda)$ for which $\mathcal{O}^{\epsilon}(\lambda)$ has edge-triangle agreement is called called the $\epsilon$-tolerance.
\end{definition}

\begin{definition}[Algorithm Mappings]
Let 
\begin{enumerate}[label=(\roman*)]
\item
$\HH_{E \rightarrow T}(\lambda)$ be the Lagrange multipliers that result from executing lines~\ref{alg:edge-to-triangle-start}-\ref{alg:edge-to-triangle-end} in Algorithm~\ref{alg:parallel-message-passing},
\item
$\HH_{T \rightarrow E}(\lambda)$ be the Lagrange multipliers that result from executing lines~\ref{alg:triangle-to-edge-start}-\ref{alg:triangle-to-edge-end} in Algorithm~\ref{alg:parallel-message-passing},
\item
$\HH = \HH_{T \rightarrow E} \circ \HH_{E \rightarrow T}$ be one pass of Algorithm~\ref{alg:parallel-message-passing},
\item
$\HH^i(\cdot) = \underbrace{\HH(\HH(\ldots(\HH(\cdot))\ldots))}_{i \text{ times}}$ be the $i$-fold composition of $\HH$. 
\end{enumerate}
\end{definition}

Note that $\HH_{E \rightarrow T}$ and $\HH_{T \rightarrow E}$ and consequently also $\HH$ are well-defined mappings since, even though Algorithm~\ref{alg:parallel-message-passing} is parallel, the update steps do not depend on the order in which they are processed.

\begin{lemma}
\label{lemma:e->t}
Let $\alpha \in (0,1]$ and let $\lambda$ be Lagrange multipliers.
Let $e \in E$ and $t \in T$ with $e \subsetneq t$.
Define new Lagrange multipliers as
\begin{equation}
\lambda'_{t',e'} = \begin{cases} 
\lambda_{t',e'} - \alpha c^{\lambda}_e, & e = e', t = t' \\
\lambda_{t',e'} , & e \neq e' \text{ or } t \neq t'
\end{cases}
\end{equation}
\begin{enumerate}[label=(\roman*)]
\item
$LB(c^{\lambda}) \leq LB(c^{\lambda'})$.
\item
$\OO_e(c^{\lambda}) \subseteq \OO_e(c^{\lambda'})$.
\item
\label{lemma:e->t:locally-optimal-intersection}
$LB(c^{\lambda}) < LB(c^{\lambda'})
\Leftrightarrow
\OO_e(c^{\lambda}) \cap \Pi_{t,e}(\OO_t(c^{\lambda'})) = \varnothing$.
\item
$LB(c^{\lambda}) = LB(c^{\lambda'}) \Rightarrow \OO_t(c^{\lambda'}) \subseteq \OO_t(c^{\lambda})$.
\item
$LB(c^{\lambda}) = LB(c^{\lambda'})$ and $c^{\lambda}_e \neq 0$
$\Rightarrow$
$\Pi_{t,e}(\OO_t(c^{\lambda'})) = \OO_e(c^{\lambda})$.
\end{enumerate}
\end{lemma}
\begin{proof}
\begin{enumerate}[label=(\roman*)]
\item
If $c^{\lambda}_e \geq 0$ then $LB(c^{\lambda})_e = LB(c^{\lambda'})_e$ and $LB(c^{\lambda})_t \leq LB(c^{\lambda'})_t$ since $c^{\lambda}_{t,e} \leq c^{\lambda'}_{t,e}$.

If $c^{\lambda}_e < 0$ then $LB(c^{\lambda})_e = c^{\lambda}_e < (1-\alpha) c^{\lambda}_e = LB(c^{\lambda'})_e$.
$\leq LB(c^{\lambda'})_t = \min_{y \in \M_{T}} c^{\lambda'}(e) \geq \min_{y \in \M_{T}} c^{\lambda}(e) - \alpha c^{\lambda}_e = LB(c^{\lambda})_t - \alpha c^{\lambda}_e$.
\item
It holds that $c^{\lambda'}_e = (1-\alpha) c^{\lambda}_e$.
Hence, if $\alpha = 1$ then $\OO_e(c^{\lambda}) = \{0,1\}$ and the claim trivially holds.
Otherwise $\OO_e(c^{\lambda}) = \OO_e(c^{\lambda})$.
\item
Assume $\OO(c^{\lambda})_t \cap \Pi_e(\OO(c^{\lambda'}_t) = \varnothing$.
Assume first that $\alpha = 1$.
Then it must hold that $\abs{\OO(c^{\lambda})_t} = 1$.
Let $\{y_e^*\} = \OO(c^{\lambda})_e$ and $y^*_t \in \argmin_{y \in \M_{T}} c^{\lambda}_t(y)$.
Let $y'_e \in \argmin_{y \in \{0,1\}} c^{\lambda'}_e y$ and $y'_t \in \argmin_{y \in \M_{T}} c^{\lambda'}_t(y)$ such that $y'_e = \Pi_e(y'_t)$ (this is possible due to $\OO(c^{\lambda'})_e = \{0,1\}$ for $\alpha = 1$.
Then
\begin{multline}
LB(c^{\lambda})_e + LB(c^{\lambda})_t = c^{\lambda}_e y_e^* + c^{\lambda}_t(y^*_e) \\
< c^{\lambda}_e y'_e + c^{\lambda}_t(y'_e) \\
= c^{\lambda'}_e y'_e + c^{\lambda'}_t(y'_e) = LB(c^{\lambda'})_e + LB(c^{\lambda'})\,.
\end{multline}
For $\alpha < 1$ the result follows from the above and the concavity of $LB$.

Assume now $\OO(c^{\lambda})_t \cap \Pi_e(\OO(c^{\lambda'}_t) \neq \varnothing$.
Choose $y_e^* \in \OO(c^{\lambda})_e$ and $y_t^* \in \OO(c^{\lambda})_t$ such that $y^*_t(e) = y_e^*$.
Then it holds that
\begin{multline}
LB(c^{\lambda})_e + LB(c^{\lambda})_t = c^{\lambda}_e y_e^* + c^{\lambda}_t(y^*_t) \\
= c^{\lambda'}_e y_e^* + c^{\lambda'}_t(y^*_t) > LB(c^{\lambda'})_e + LB(c^{\lambda'})_t
\end{multline}
Since $LB$ is non-decreasing, it follows that $LB(c^{\lambda}) = LB(c^{\lambda'})$.

\item
If $c^{\lambda}_e = 0$ there is nothing to show since $\lambda' = \lambda$.

Assume that $c^{\lambda}_e > 0$.
Then it must hold that $0 \in \Pi_{t,e}(\OO_t(c^{\lambda}))$ due to~\ref{lemma:e->t:locally-optimal-intersection}.
Since $c^{\lambda'}_{t}(e) > c^{\lambda}_{t}(e)$ and all other costs stay the same, it holds that
\begin{equation}
\label{eq:e->t-case-by-case-optimal-set}
    y_t \begin{cases}
    \in \OO_t(c^{\lambda'}), & y_t \in \OO_t(c^{\lambda}), y_t(e) = 0 \\
    \notin \OO_t(c^{\lambda'}), & y_t \notin \OO_t(c^{\lambda}), y_t(e) = 0 \\
    \notin \OO_t(c^{\lambda'}), & y_t \in \OO_t(c^{\lambda}), y_t(e) = 1 \\
    \notin \OO_t(c^{\lambda'}), & y_t \notin \OO_t(c^{\lambda}), y_t(e) = 1 
    \end{cases}\,.
\end{equation}
Hence, the result follows.

The case $c^{\lambda}_e < 0$ can be proved analoguously.
\item Follows from the case by case analysis in~\eqref{eq:e->t-case-by-case-optimal-set}
\end{enumerate}
\end{proof}

\begin{lemma}
\label{lemma:t->e}
Let $\alpha \in (0,1]$ and let $\lambda$ be Lagrange multipliers.
Let $e \in E$ and $t \in T$ with $e \subsetneq t$.
Define
\begin{equation}
\lambda'_{t',e'} = \begin{cases} 
\lambda_{t',e'} + \alpha m_{t \rightarrow e}(c^{\lambda}_t), & e = e', t = t' \\
\lambda_{t',e'} , & e \neq e' \text{ or } t \neq t'
\end{cases}
\end{equation}
\begin{enumerate}[label=(\roman*)]
\item
$LB(c^{\lambda}) \leq LB(c^{\lambda'})$.
\item
$\OO_t(c^{\lambda}) \subseteq \OO_t(c^{\lambda'})$.
\item
$LB(c^{\lambda}) < LB(c^{\lambda'})
\Leftrightarrow
\OO_e(c^{\lambda}) \neq \Pi_{t,e}(\OO_t(c^{\lambda'})$.
\item
$
LB(c^{\lambda}) = LB(c^{\lambda'}) 
\Rightarrow
\OO_e(c^{\lambda'}) \subseteq \OO_e(c^{\lambda})
$
\item
$LB(c^{\lambda}) = LB(c^{\lambda'})$ and
$m_{t \rightarrow e}(c^{\lambda}) \neq 0$
$\Rightarrow$
$\Pi_{t,e}(\OO_t(c^{\lambda})) = \OO_e(c^{\lambda'})$.
\end{enumerate}
\end{lemma}
\begin{proof}
Analoguous to the proof of Lemma~\ref{lemma:e->t}.
\end{proof}

\begin{lemma}
\label{lemma:lb-monotonicity}
Each iteration of Algorithm~\ref{alg:parallel-message-passing} is non-decreasing in the lower bound $LB$ from~\eqref{eq:dual-multicut}.
\end{lemma}
\begin{proof}
Follows from Lemma~\ref{lemma:e->t}~(i) and Lemma~\ref{lemma:t->e}~(i).
\end{proof}

\begin{lemma}
\label{lemma:locally-optimal-non-increasing-e}
If $LB(c^{\lambda}) = LB(\HH(c^{\lambda}))$ then
$\OO_e(\HH(c^{\lambda})) \subseteq \OO_e(c^{\lambda})$ for all $e \in E$.
\end{lemma}
\begin{proof}
If $\OO_e(c^{\lambda}) = \{0,1\}$, there is nothing to show.

Assume $\{0\} = \OO_e(c^{\lambda})$.
Then $\Pi_{t,e}(\HHet(c^{\lambda})_t) = \{0\}$ due to Lemma~\ref{lemma:e->t} (iv) for all $t \in T$, $e \subsetneq t$.
Then Lemma~\ref{lemma:t->e} (v) implies that $\OO_e(\HH(c^{\lambda})) = \{0\}$.

The case $\{1\} = \OO_e(c^{\lambda})$ can be proved analoguously. 
\end{proof}

\begin{lemma}
\label{lemma:locally-optimal-non-increasing-t}
If $LB(c^{\lambda}) = LB(\HHet \circ \HHte(c^{\lambda}))$ then
$\Pi_{t,e}(\OO_t(\HHet \circ \HHte(c^{\lambda}))) \subseteq \Pi_{t,e}(\OO_t(c^{\lambda}))$ for all $t \in T$, $e \in E$ and $e \subsetneq t$.
\end{lemma}
\begin{proof}
Write $c^{\lambda'} = \HHte(c^{\lambda})$ and $c^{\lambda''} = \HHet(c^{\lambda'})$.
Let some $t \in T$, $e \in E$ and $e \subsetneq t$ be given.
If $m_{t\rightarrow e} = 0$ the result follows from Lemma~\ref{lemma:e->t}~(iv).
Hence we can assume that $m_{t\rightarrow e} \neq 0$.
Lemma~\ref{lemma:t->e}~(iii) and~(v) imply $\Pi_{t,e}(\OO_t(c^{\lambda})) = \OO_e(c^{\lambda'})$.
Due to Lemma~\ref{lemma:e->t}~(iv) the result follows.
\end{proof}

\begin{lemma}
\label{lemma:locally-optimal-shrinking}
Define 
$\xi_e = \OO_e(c^{\lambda})$ for all $e \in E$,
$\xi_t = \OO_t(\HHet(c^{\lambda}))$,
$\xi'_e = \OO_e(\HH(c^{\lambda}))$ for all $e \in E$ and
$\xi'_t = \OO_t(\HHet\circ\HH(c^{\lambda}))$.
If $LB(c^{\lambda}) = LB(\HHet\circ\HH(c^{\lambda}))$ and $\xi$ is not arc-consistent
then $\exists e \in E$ such that $\xi'_e \subsetneq \xi_e$ or $\exists$ $t \in T$, $e\in E$ and $e\subsetneq t$ such that $\Pi_{t,e}(\xi'_t) \subsetneq \Pi_{t,e}(\xi_t)$.
\end{lemma}
\begin{proof}
If $\xi$ is not arc-consistent there exists $e \in E$, $t \in T$ and $e \subsetneq t$ such that $\xi_e \neq \Pi_{t,e}(\xi_t)$.

The case $\abs{\xi_e} = 1 = \abs{\Pi_{t,e}(\xi_t)}$ implies $\xi_e \cap \Pi_{t,e}(\xi_t) = \varnothing$ and due to Lemma~\ref{lemma:t->e}~(iii) contradicts that $LB$ is not increasing.

Assume $\xi_e \subsetneq \Pi_{t,e}(\xi_t)$.
Due to Lemma~\ref{lemma:e->t}~(iv) and (v) this would imply an increase in the lower bound.

Hence we can assume that $\Pi_{t,e}(\xi_t) \subsetneq \xi_e$.
Due to Lemma~\ref{lemma:t->e}~(v) it holds that $\abs{\xi'_e} = 1$.

Together with Lemmas~\ref{lemma:locally-optimal-non-increasing-e} and~\ref{lemma:locally-optimal-non-increasing-t} the result follows.
\end{proof}

\begin{lemma}
$\HH$ is a continuous mapping.
\end{lemma}
\begin{proof}
All the operations used in Algorithm~\ref{alg:parallel-message-passing} are continuous, i.e.\ adding and subtracting, dividing by a constant and taking the minimum w.r.t.\ elements for the min-marginals.
Hence, $\HH$, the composition all such continuous operations, is continuous again.
\end{proof}

\begin{lemma}
The lower bound $LB$ from~\eqref{eq:dual-multicut} is continuous in $\lambda$.
\end{lemma}
\begin{proof}
Taking minima is continuous as well as addition. Hence $LB$ is continuous as well.
\end{proof}

\begin{lemma}
$\epsilon$-tolerance is continuous in $\lambda$.
\end{lemma}
\begin{proof}
We first prove that for any arc-consistent subset $\xi$ the minimal $\epsilon$ for which $\xi \subseteq \OO^{\epsilon}(\lambda)$ is continuous.
To this end, note that the minimum $\epsilon$ such that $\xi_e \subseteq \OO^{\epsilon}_e(\lambda)$ for any edge $e \in E$ can be computed as  
\begin{equation}
\xi_e = \begin{cases}
c^{\lambda}_e,& \xi = \{0\} \\
-c^{\lambda}_e,& \xi = \{1\} \\
\abs{c^{\lambda}_e},& \xi = \{0,0\}
\end{cases}
\end{equation}
All the expressions are continuous, hence the minimum $\epsilon$ for any edge is continuous. A similar observation holds for triangles.
Since the $\xi$-specific $\epsilon$ is the maximum over all edges and triangles, it is continuous as well.

Since the $\epsilon$-tolerance is the minimum over all minimal $\xi$-specific $\epsilon$ and there is a finite number of arc-consistent subsets $\xi$, the result follows.
\end{proof}

\begin{lemma}
For any edge costs $c \in \R^{E}$ there exists $M > 0$ such that $\norm{\HH^i(c)} \leq M$ for any $i \in \N$.
\end{lemma}
\begin{proof}
Assume $\HH^i(c)$ is unbounded.
If all $\HH^i(c)_t$ $t \in T$ are bounded, all $\HH^i(c)_e$ are bounded as well due to~\eqref{eq:reparametrization}.
Hence, there must exist $t \in T$ such that $\HH^i(c)_t$ is unbounded.
Since $LB(H^i)(c)_t$ is bounded below by Lemma~\ref{lemma:lb-monotonicity} and trivially above by $0$, it must hold that either
\begin{enumerate}[label=(\roman*)]
\item there exists one edge $e \subsetneq t$ such that  $\HH^i(c)_t(e)$ converges towards $-\infty$ on a subsequence or
\item there exists at most one edge $e \subsetneq t$ such that $\HH^i(c)_t(e)$ converges towards $\infty$ and there exist $e'\neq e'' \subsetneq t$ with $e' \neq e$ and $e'' \neq e$ such that $\HH^i(c)_t(e')$ and $\HH^i(c)_t(e'')$ converge towards $-\infty$ with
$\HH^i(c)_t(e) - \HH^i(c)_t(e') \leq M'$ and
$\HH^i(c)_t(e) - \HH^i(c)_t(e'') \leq M'$ where $M' > 0$ is a constant, since otherwise $LB(\HH^i(c))_t$ would converge to $-\infty$.
\end{enumerate}
Hence there must be at least double the number of Lagrange multipliers $\lambda_{t,e}$ that converge towards $-\infty$ than those that converge towards $\infty$ with at least the same rate.
Hence, there must be $\tilde{e} \in E$ such that on a subsequence $\HH^i(c)_{\tilde{e}}$ converges towards $-\infty$, contadicting that $LB(\HH^i(c))_e$ is bounded below by Lemma~\ref{lemma:lb-monotonicity}.
\end{proof}

\begin{proof}[Proof of Theorem~\ref{thm:edge-triangle-agreement-convergence}]
Due to the Bolzano Weierstrass theorem and the boundedness of $\HH^i(c)$ there exists a subsequence $i(k)$ such that $\HH^{i(k)}(c)$ converges to a $c^{\lambda^*}$.
We first show that $\epsilon(c^{\lambda^*}) = 0$.
Since $\HH$ and $LB$ are continuous an $LB$ is non-decreasing, we have 
\begin{multline}
    LB(c^{\lambda^*}) = \lim_{k \rightarrow \infty} LB(\HH^{i(k)}(c)) \\
    = \lim_{k \rightarrow \infty} LB(\HH^{i(k) + n}(c)) \quad \forall n \geq 0\,.
\end{multline}
Due to Lemma~\ref{lemma:locally-optimal-shrinking} and $\epsilon$ being continuous, $\epsilon(c^{\lambda^*}) = 0$ follows.

Define $s^i = \max_{j \leq i} \epsilon(\HH^j(c))$.
Then $s^i$ is by construction a non-negative non-decreasing sequence and therefore has a limit $s^*$.
Hence, there also must exist a subsequence $j(k)$ such that $\lim_{k \rightarrow \infty}\epsilon(\HH^{j(k)}(c)) = s^*$.
As proved above the subsequence $j(k)$ has a subsequence which converges towards $\epsilon(\cdot) = 0$, hence $s^* = 0$ as well.
Finally, 
\begin{equation}
0 \leq \epsilon(\HH^i(c)) \leq s^i
\end{equation}
implies convergence towards node-triangle agreement.
\end{proof}

\subsection{GPU implementations}
\paragraph{Edge contraction}
We use a specialized implementation for edge contraction using Thrust~\cite{Thrust} which is faster than performing it via general sparse matrix-matrix multiplication routines and most importantly has lesser memory footprint allowing to run larger instances. We store the adjacency matrix $A = (I, J, C)$ in COO format, where $I, J, C$ correspond to row indices, column indices and edge costs resp. The pseudocode is given in Algorithm~\ref{alg:gpu-edge-contraction}.

\begin{algorithm}
\DontPrintSemicolon
\caption{\texttt{GPU Edge-Contraction}}
\label{alg:gpu-edge-contraction}
\KwData{Adjacency matrix $A = (I, J, C)$, Contraction mapping $f : V \rightarrow V'$}
\KwResult{Contracted adjacency matrix $A' = (I', J', C')$}
\tcp{Assign new node IDs}
$\hat{I}(v) = I(f(v)),\ \forall v \in V$ \;
$\hat{J}(v) = J(f(v)),\ \forall v \in V$ \;
\texttt{COO-Sorting}($\hat{I}, \hat{J}, C$)\;
\tcp{Remove duplicates and add costs}
$(I', J', C') =$ \texttt{reduce\_by\_key}(\;
\nonl\Indp \texttt{keys}$ = (\hat{I}, \hat{J}), $\texttt{values}$ = \hat{C}, $ \texttt{acc} $= +)$ \;
\end{algorithm}

\paragraph{Conflicted cycles}
For detecting conflicted cycles we use specialized CUDA kernels. The pseudocode for detecting 5-cycles is given in Algorithm~\ref{alg:gpu-5-cycles}. The algorithm searches for conflicted cycles in parallel in the positive neighbourhood $\NE^+$ of each negative edge. 
To efficiently check for intersection in Line~\ref{alg:check-intersect-5-cycles} we store the adjacency matrix in CSR format. 
\begin{algorithm}
\DontPrintSemicolon
\caption{\texttt{Parallel Conf.\ 5-Cycles}}
\label{alg:gpu-5-cycles}
\KwData{Adjacency matrix $A=(V,E,c)$}
\KwResult{Conflicted cycles $Y$ in $A$}
\tcp{Partition edges based on costs}
$E^+ = \{ij \in E: c_{ij} > 0\}$\;
$E^- = \{ij \in E: c_{ij} < 0\}$\;
$Y = \varnothing $ \;
\tcp{Check for attractive paths}
\For{$v_1v_3 \in \NE^+(v_0) \times \NE^+(v_4): v_0v_4 \in E^-$ in parallel}{
    \label{alg:check-intersect-5-cycles}
    \For{$v_2 \in \NE^+(v_1) \cap \NE^+(v_3)$}{
        $Y = Y \cup \{v_0, v_1, v_2, v_3, v_4 \}$ \;
    }
}
\end{algorithm}
\subsection{Results comparison}
\begin{figure*}
  \begin{subfigure}[b]{0.5\linewidth}
    \centering
    \includegraphics[width=0.9\linewidth]{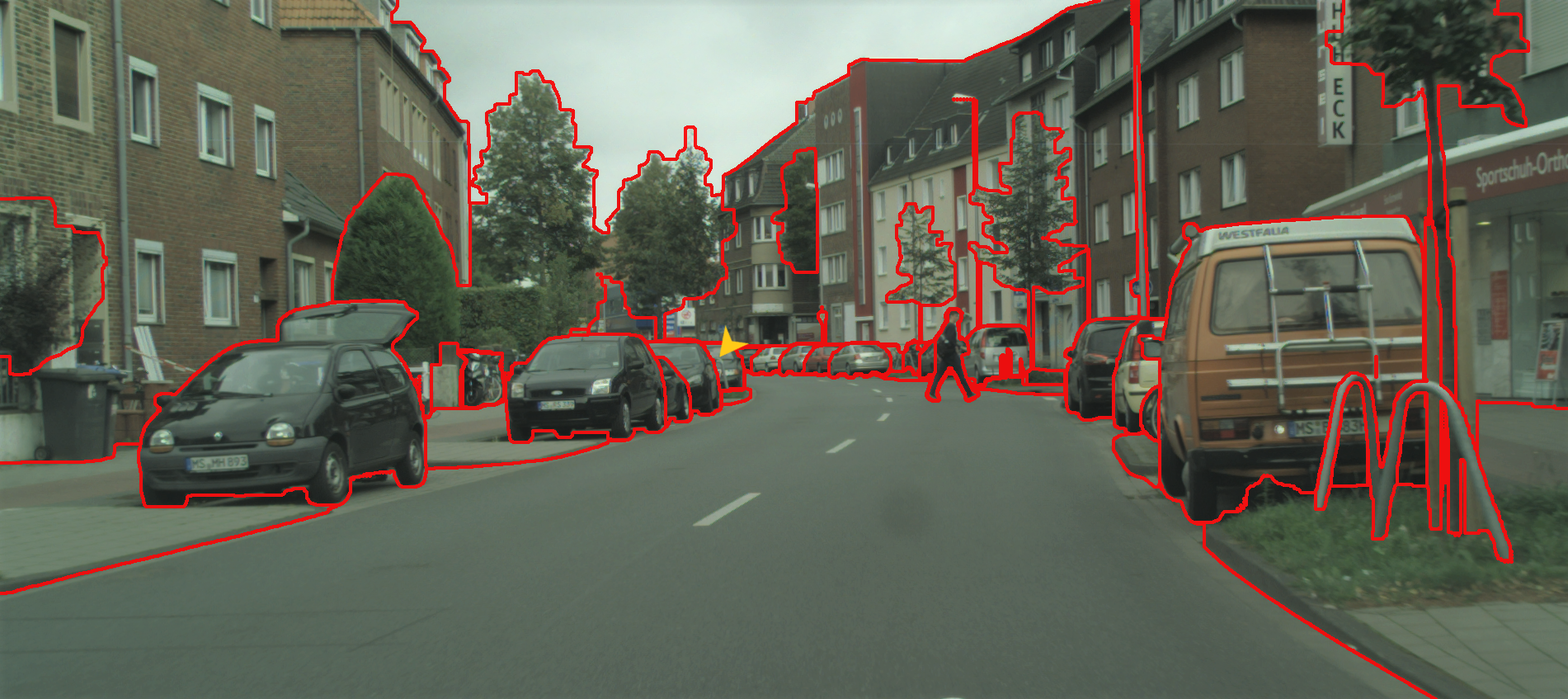} 
    \caption{\texttt{GAEC}~\cite{keuper2015efficient}, Cost = -2455070, time: $12.8$s} 
    \label{fig7:a} 
    \vspace{4ex}
  \end{subfigure}
  \begin{subfigure}[b]{0.5\linewidth}
    \centering
    \includegraphics[width=0.9\linewidth]{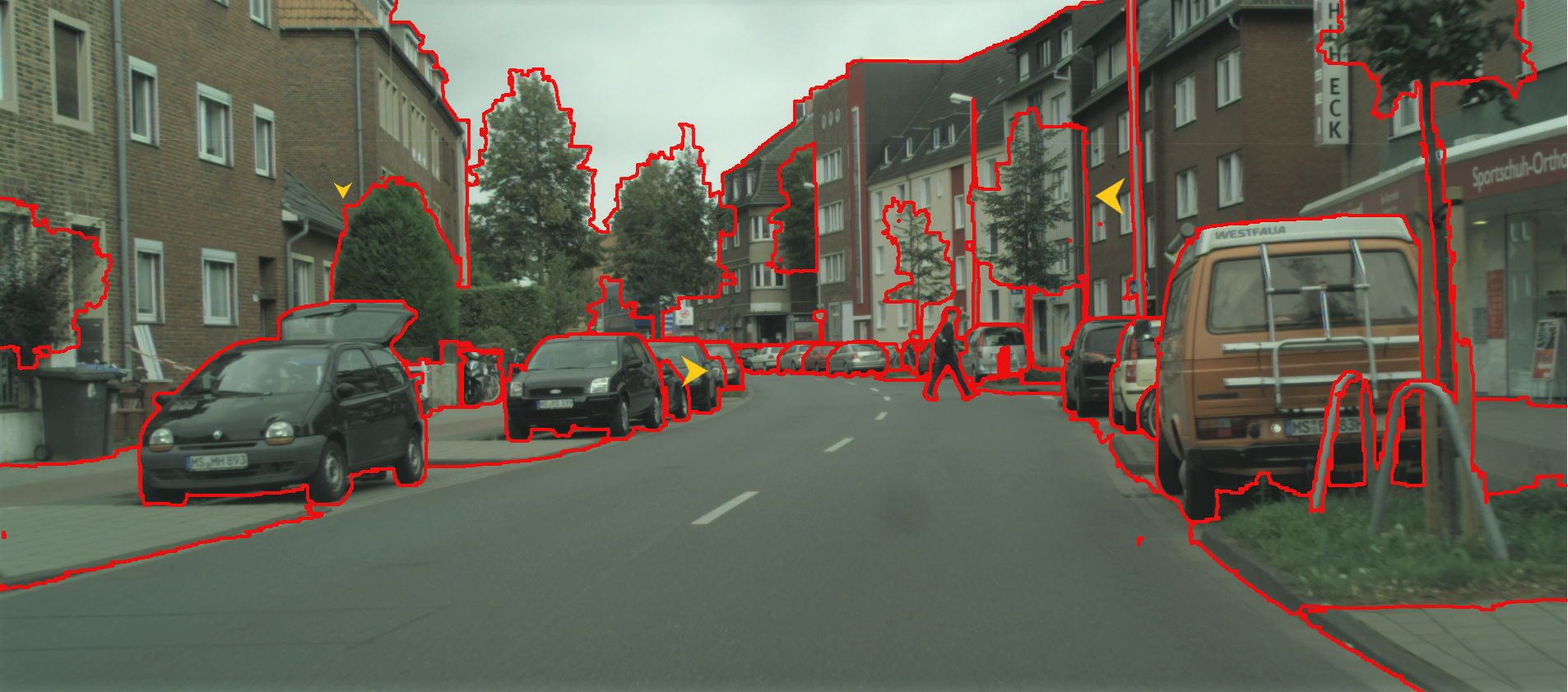} 
    \caption{P, cost = -2347254, time: $\textbf{0.4s}$ } 
    \label{fig7:b} 
    \vspace{4ex}
  \end{subfigure} 
  \begin{subfigure}[b]{0.5\linewidth}
    \centering
    \includegraphics[width=0.9\linewidth]{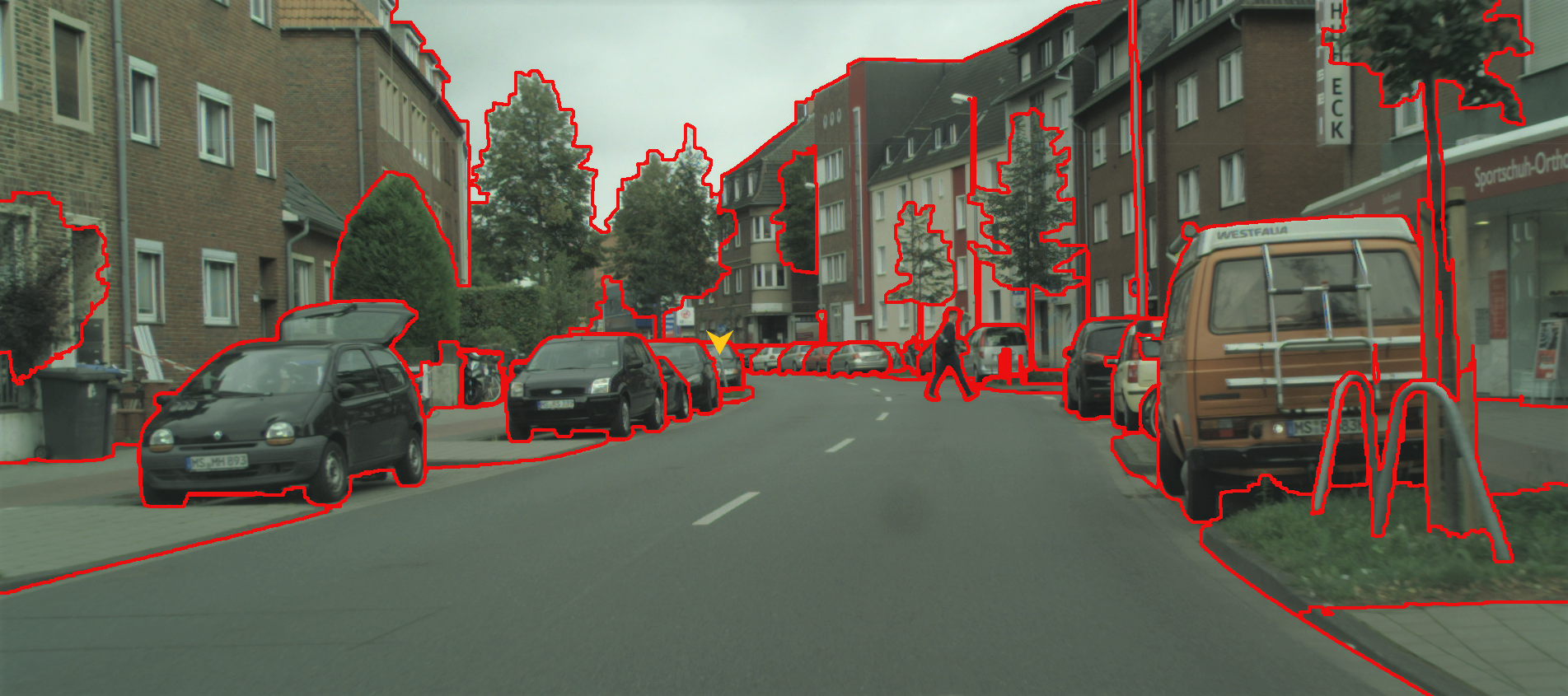} 
    \caption{\texttt{PD}, cost = -2499152, time: $1.1$s }
    \label{fig7:c} 
  \end{subfigure}
  \begin{subfigure}[b]{0.5\linewidth}
    \centering
    \includegraphics[width=0.9\linewidth]{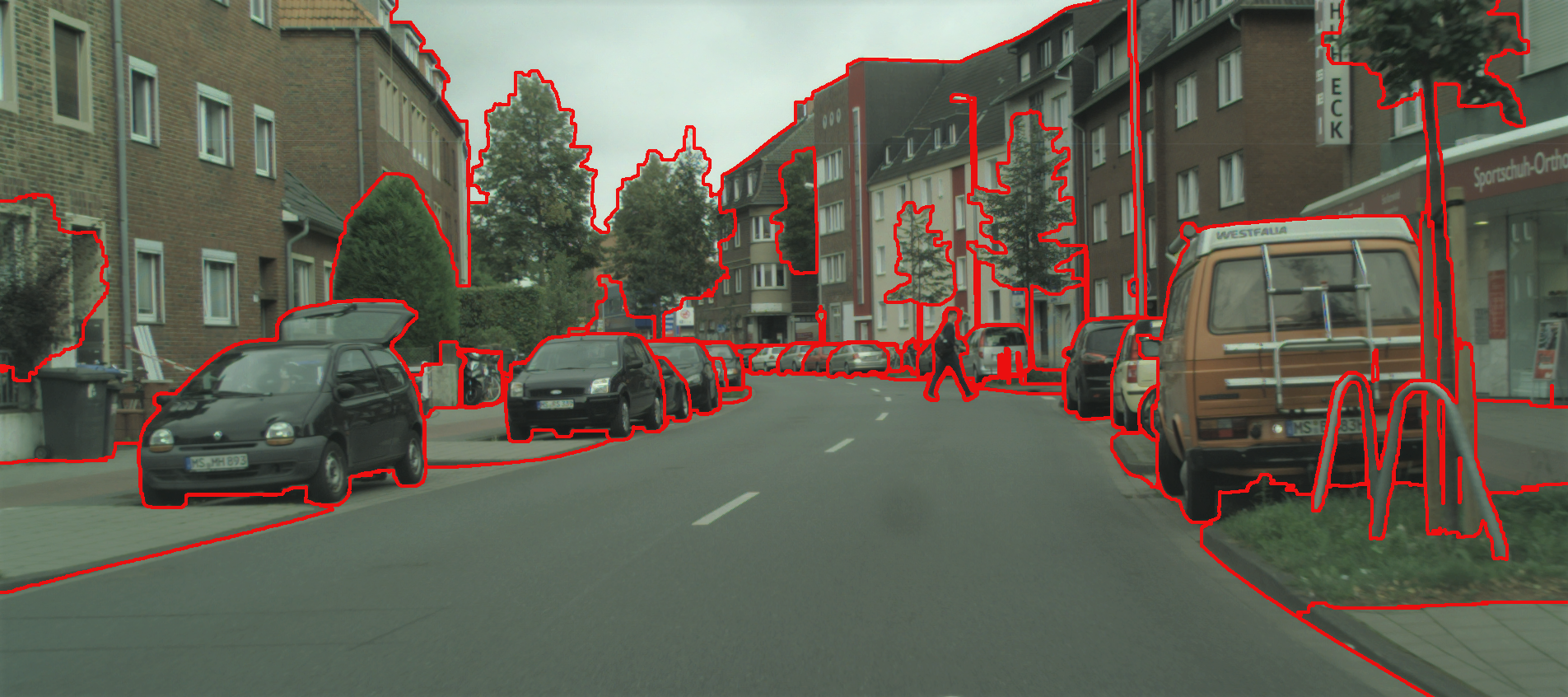} 
    \caption{\texttt{PD+}, cost = $-\textbf{2523547}$, time: $2.2$s}
    \label{fig7:d} 
  \end{subfigure} 
  \caption{Results comparison on an instance of \normalfont\textit{Cityscapes} dataset highlighting the transitions. Yellow arrows indicate incorrect regions. Our purely primal algorithm (\texttt{P}) suffers in localizing the sidewalks and trees. \texttt{PD+} is able to detect an occluded car on the left side of the road which all other methods did not detect. (Best viewed digitally)}
  \label{fig:cityscapes_results_comparison} 
\end{figure*}

\end{document}